\documentclass[twocolumn]{aastex63}

\usepackage{soul}
\usepackage{color}
\usepackage{hyperref}
\usepackage{graphicx}
\usepackage[percent]{overpic}

\usepackage{mathtools}
\usepackage{enumerate}  
\usepackage{ulem}
\usepackage{amsmath}
\usepackage{rotating}

\newcommand{\om}{$\omega$}
\newcommand\eg[1]{\textcolor{blue}{ #1}}

\newcommand{\nukol}{\nu_{\rm{Kol}}}
\newcommand{\nukra}{\ensuremath{\nu_{\rm{Kra}}}}
\newcommand{\omegacut}{\ensuremath{\omega_{\rm{cut}}}}
\newcommand{\ebreak}{\ensuremath{E_{\rm{break}}}}
\newcommand{\ecut}{\ensuremath{E_{\rm{cut}}}}

\shorttitle{Jitter radiation in SNRs}
\shortauthors{Greco et al.}

\begin{document}

\title{Jitter radiation as an alternative mechanism for the nonthermal X-ray emission of Cassiopeia A}

\correspondingauthor{Emanuele Greco}
\email{e.greco@uva.nl}

\author[0000-0001-5792-0690]{Emanuele Greco}
\affiliation{Anton Pannekoek Institute for Astronomy \& GRAPPA, University of Amsterdam, Science Park 904, 1098 XH Amsterdam, The Netherlands}

\author[0000-0002-4708-4219]{Jacco Vink}
\affiliation{Anton Pannekoek Institute for Astronomy \& GRAPPA, University of Amsterdam, Science Park 904, 1098 XH Amsterdam, The Netherlands}

\author[0000-0002-1038-3370]{Amael Ellien}
\affiliation{Anton Pannekoek Institute for Astronomy \& GRAPPA, University of Amsterdam, Science Park 904, 1098 XH Amsterdam, The Netherlands}

\author[0000-0003-1429-1059]{Carlo Ferrigno}
\affiliation{Department of Astronomy, University of Geneva, Chemin d’Ecogia 16, CH-1290 Versoix, Switzerland}
\affiliation{INAF, Osservatorio Astronomico di Brera, Via E. Bianchi 46, I-23807, Merate, Italy}

\begin{abstract}
Synchrotron radiation from relativistic electrons is usually invoked as the responsible for the nonthermal emission observed in Supernova Remnants (SNRs). Diffusive shock acceleration (DSA) is the most popular mechanism 
to explain the process of particles acceleration and within its framework a crucial role is played by the turbulent magnetic-field. However, the standard models commonly used to fit X-ray synchrotron emission do not take into account the effects of turbulence in the shape of the resulting photon spectra. An alternative mechanism that properly includes such effects is the jitter radiation, that provides for an additional power-law beyond the classical synchrotron cutoff. We fitted a jitter spectral model to Chandra, NuSTAR, SWIFT/BAT and INTEGRAL/ISGRI spectra of Cassiopeia A and found that it describes the X-ray soft-to-hard range better than any of the standard cutoff models. The jitter radiation allows us to measure the index of the magnetic turbulence spectrum $\nu_B$ and the minimum scale of the turbulence $\lambda_{\rm{min}}$ across several regions of Cas A, with best-fit values $\nu_B \sim 2-2.4$ and $\lambda_{\rm{min}} \lesssim 100$ km. 
\end{abstract}

\section{Introduction}
\label{sect:intro}


It is generally accepted that particle acceleration at the shock fronts of supernova remnants (SNRs) is the result of diffusive shock acceleration (DSA) \citep{bel78,md01}.
According to DSA, charged particles repeatedly cross the shock front back and forth due to the magnetic-field irregularities, gaining a few percent of energy with each crossing. 
%
Among the accelerated particles are highly relativistic electrons, and those with energies $\gtrsim 10$~TeV will produce
X-ray synchrotron radiation commonly observed in young SNRs (see \citealt{vin20} for a review). To reach those very high energies, the electrons need to reside close to the shock front, enabling fast repeated shock crossings.  For this confinement close to the shock front a highly turbulent magnetic field near the shock front is necessary.

Magnetic turbulence greatly affects the polarization of the light emitted. Synchrotron radiation 
from a nonthermal population of electrons with a power-law index of energy of 2--3 is intrinsically polarized at the 70--75\%
level (see Eq. 3.29 in \citealt{gs65}), but Cassiopeia A (Cas A) observations performed by the Imaging X-ray Polarimetry Explorer (IXPE) (\citealt{IXPE}), showed that the degree polarization of the radiation in the 4-6 keV energy band is roughly $3\%$ (\citealt{vpf22}), suggesting a highly turbulent magnetic field.

Despite the crucial need for high turbulence to achieve nonthermal X-ray emitting electrons and the recent X-ray polarization results, the most common models used to fit X-ray synchrotron spectra do not take into account turbulence effect in shaping the resulting photon spectra. The synchrotron spectral shape emitted by a population of electrons having a power-law distribution with high-energy cutoff with particle index $\xi$ is, in the assumption of uniform magnetic field:

\begin{equation}
    n(\hbar\omega) \propto (\hbar\omega)^{-\Gamma} \rm{exp} \bigg[ - \Big( \frac{\hbar\omega}{\hbar\omega_{c}}\Big)^{\beta} \bigg],
\end{equation}

where $\Gamma = (\xi+1)/2$. In the loss-limited scenario adopted in \citet{za07}, $\beta = 0.5$ while $\beta \approx 1$ in the \texttt{srcut} model available in XSPEC \citet{rk99}. The latter has been the most commonly used to fit X-ray synchrotron-dominated spectra of SNR--- e.g. \citealt{byy05,rbg08,mbd13}---but recently 
X-ray SNR spectra were also analyzed using the relations from \citet{za07}, see e.g. \cite{tuk21}, \cite{gmc22} and \cite{smb22}. In particular, \citet{za07} derived the analytical expression for the spectra of shock-accelerated electrons in the scenario in which the electron energy is loss-limited. They also recovered the non-thermal synchrotron spectra emitted by a population of such electrons. 
It is also common to describe the synchrotron emission with a simple power-law spectrum with the corresponding photon index $\Gamma \gtrsim 2.5$  (\citealt{byy05,chb07,hvb12}). 
This is much steeper than the typical radio spectral index of SNRs of $\alpha\approx 0.5$, corresponding to 
the photon index $\Gamma=\alpha +1 \approx 1.5$ or, 
for energies beyond the cooling break,
$\Gamma_{\rm cb}=\Gamma + 0.5 \approx 2$.
This discrepancy has been usually explained by taking into account the curvature of the spectrum due to the cut-off expected in both age-limited (\citealt{rk99}) and loss-limited scenarios (\citealt{za07}). However, there are several detections of power-law spectra extending well beyond the soft X-ray band (i.e. $\gtrsim 10$ keV), at odds with the continuously steepening to be expected from
a spectrum beyond the spectral cutoff.

The best example is provided by the SNR Cas A, a young ($\sim$ 350 years old) and bright core-collapse SNR located at a distance of 3.4 kpc (\citealt{rhf95}). 
Cas A shows a power-law extending at least up to 50 keV in NuSTAR data (\citealt{grh15}) and up to 100 keV according to RXTE data (\citealt{akg97}), BeppoSAX \citep{vl03} and more recent publications based on INTEGRAL and SWIFT (\citealt{rvd06,wz16}).

\begin{figure}[!ht]
    \centering
    \includegraphics[width=.99\columnwidth]{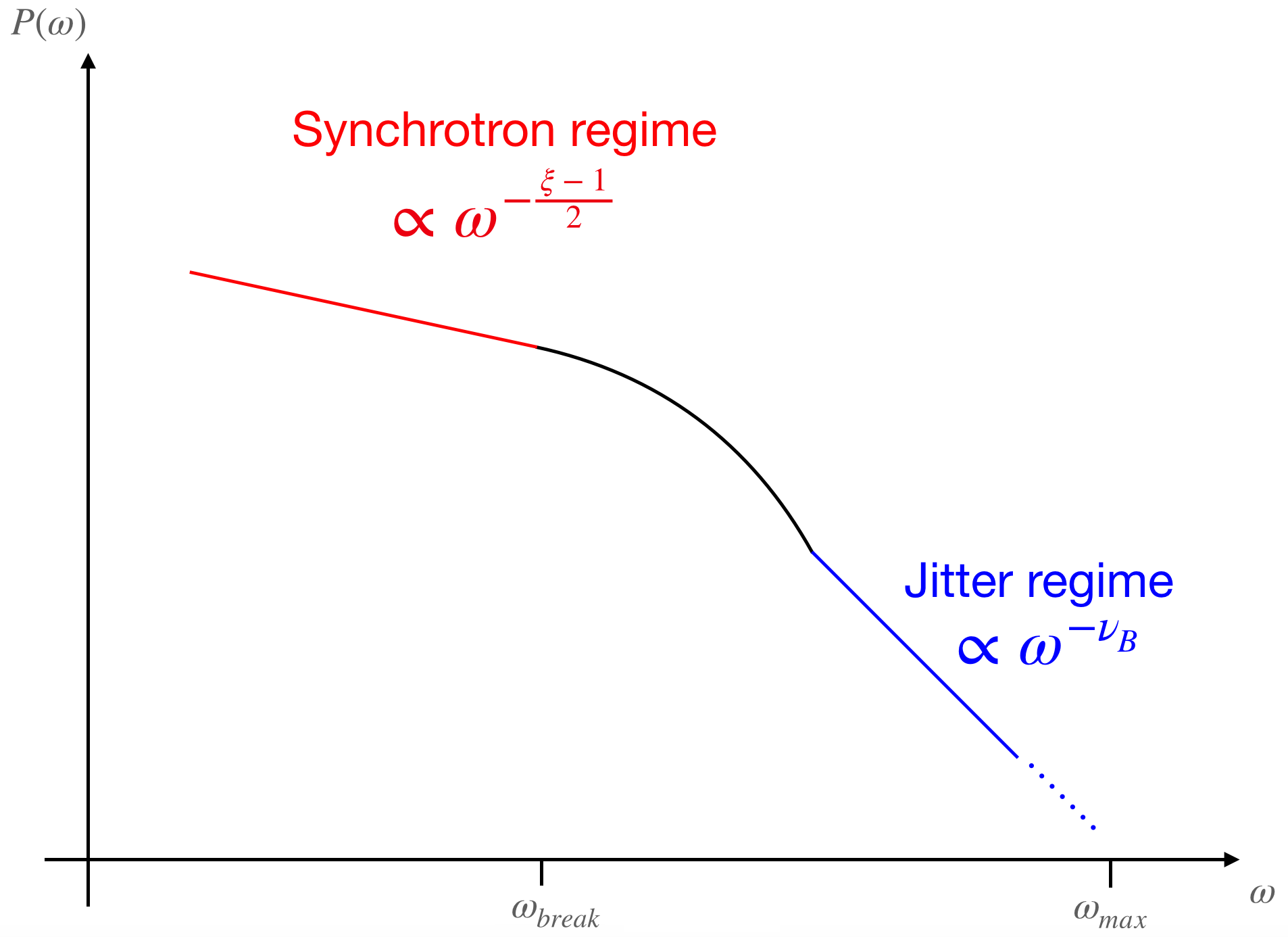}
    \caption{Radiation spectrum $P(\omega)$ produced by an ensemble of electron in a turbulent magnetic field (adapted from TF87). 
    See also \citet{rk10} (their Fig. 5 and 6) for numerical results on the single particle emissivity.
    See text for details.}
    \label{fig:toptygyn}
\end{figure}

It is therefore natural to ask if a different mechanism, potentially linked to the often overlooked small-scale magnetic-field turbulence, might be responsible for this extended power-law component.  The spectrum resulting from a motion of a relativistic electron in a highly non-uniform magnetic field was first investigated by \citet{tf87} (hereafter TF87) and, more recently, by \citet{kak13} (hereafter K13). The analytical results have also been numerically verified by \citet{rk10}, but for the emissivity of a single particle.
These authors refer to this radiation as \emph{jitter radiation}. They show that an ensemble of relativistic electrons embedded in a highly turbulent magnetic field leads to a 
more sophisticated spectrum than the standard synchrotron one (see Fig. \ref{fig:toptygyn}, adapted from TF87).

Jitter radiation is sensitive to turbulence on length scales of 

\begin{equation}
\label{eq:lambda}
\lambda \ll 170(B/100~{\rm \mu G})^{-1} ~{\rm km},
\end{equation}
with $B=100~{\rm \mu G}$--$500~{\rm \mu G}$ being typical for the strengths reported for Cas A \citep{hvb12}. These are scales much smaller than usually considered for magnetic-field turbulence responsible for cosmic-ray acceleration ($\sim 10^{14}-10^{15}$ cm). The main factor limiting the magnetic turbulence at short scale is Landau damping   \citet{} which is relevant at a scale $\lambda_{Ld} = \sqrt{\frac{T_e}{4\pi e^2 n_e}}$ with $T_e$ the electron temperature and $n_e$ the electron density. In order to make jitter radiation possible, it is then needed that photon formation length $\lambda_{pf} \gg \lambda_{Ld}$. Considering that a typical temperature of Cas A is 3 keV and an upstream density is $\approx 2$ cm$^{-3}$ (\citealt{hl12}) we obtain: $\lambda \gg \lambda_{Ld} \approx 0.15$ km.

From Fig. \ref{fig:toptygyn} we can distinguish three main regimes: when the frequency $\omega$ is much lower than \om$_{break}$ the radiation spectrum P($\omega$) is a power-law with an exponential $\alpha = -\frac{\xi-1}{2}$, consistent with the prediction for the synchrotron radiation in a homogeneous magnetic field (e.g. \citealt{za07}); for \om $~\approx$ \om$_{break}$ P(\om) faces a decline, whose characteristics depend on the cutoff parameter $\beta$; for \om $\gg$ \om$_{break}$, i.e. outside of this intermediate regime,  P(\om) is, again, a straight power-law but here the exponential $\nu_B$ is the exponential of the magnetic-field turbulence spectrum. 
We here do not focus on the mathematical aspects of jitter radiation (the whole derivation can be found in TF87 and K13), but we just want to highlight the following: \om$_{\rm break}$ is the break frequency of the jitter radiation, related to the cutoff frequency $\omega_{\rm cut}$ of standard synchrotron, i.e. with the electrons embedded in an homogeneous magnetic field, as $\omega_{\rm break} = \omega_{\rm cut} \times \frac{R_{\rm{L,j}}}{\lambda}$ with $R_{\rm{L,j}}$ the nonrelativistic Larmor radius; \om$_{\rm max}$ = \om$_{\rm cut} \times (R_{\rm L,j}/\lambda)^3$ is the maximum frequency to which the additional power-law extends (K13).

In this paper, we analyzed Chandra/ACIS-S, NuSTAR/FPMA,B, INTEGRAL/ISGRI and SWIFT/BAT data of Cas A, aiming at investigating the relative merits of the standard model, consisting of a synchrotron spectrum with exponential cutoff, versus its extension in the case of a highly turbulent magnetic-field, the jitter radiation model. Interestingly, the jitter model can be used to estimate  both the magnetic-field turbulence spectral index and the minimum length scale of magnetic-field turbulence.


The paper is divided as follows: in Sect. \ref{sect:x-ray_data} we present all the X-ray data considered, we describe their reduction and present results of image analysis; in Sect. \ref{sect:spec_anal} we show the results of the spectral analysis; in Sect. \ref{sect:disc} we discuss the results and comment on their implications; in Sect. \ref{sect:conc} we wrap up the main findings and draw our conclusion.

\section{X-ray data reduction and regions selection}
\label{sect:x-ray_data}

We considered X-ray observations of Cas A in order to cover a wide X-ray energy range, from 0.5 keV up to roughly 100 keV. We analyzed data collected by Chandra/ACIS-S (operating in the 0.1-10 keV, see Sect. \ref{sect:chandra}, \citealt{ACIS}), NuSTAR (3-79 keV, Sect. \ref{sect:nu}, \citealt{NuSTAR}), INTEGRAL/ISGRI (15 keV - 1 MeV, Sect. \ref{sect:isgri}, \citealt{isgri}), SWIFT/BAT (20-200 keV, Sect. \ref{sect:bat}, \citealt{swiftbat}). All the observations considered are listed in Table \ref{tab:obs}. Since INTEGRAL/ISGRI and SWIFT/BAT are coded masks and are not able to resolve Cas A, it was possible to perform spatially resolved spectral analysis only with Chandra and NuSTAR telescopes. 

\begin{table}[!ht]
    \centering
    \caption{Observation log table}
    \begin{tabular}{c|c|c|c}
      Telescope & Obs ID & PI & Exposure (Ms) \\
      \hline 
      Chandra & 4638 & Hwang & 0.164 \\ 
      \hline
         & 40001019002 & Harrison & 0.294 \\
         & 40021002002 & Harrison & 0.288 \\
         & 40021011002 & Harrison & 0.246 \\
         & 40021012002 & Harrison & 0.239 \\
         & 40021003003 & Harrison & 0.233 \\
         & 40021001005 & Harrison & 0.228 \\
      NuSTAR   & 40021002008 & Harrison & 0.226 \\
         & 40021001002 & Harrison & 0.190 \\
         & 40021015003 & Harrison & 0.160 \\
         & 40021002006 & Harrison & 0.159 \\
         & 40021015002 & Harrison & 0.86 \\
         & \multicolumn{2}{c|}{Total} & 2.3 \\ 
         \hline
        INTEGRAL & \multicolumn{2}{c|}{See Sect. \ref{sect:isgri}} & 10\\
        \hline 
        SWIFT & \multicolumn{2}{c|}{See Sect. \ref{sect:bat}} & 180 \\ 
    \end{tabular}
    \label{tab:obs}
\end{table}

\subsection{Chandra}
\label{sect:chandra}

\begin{figure*}
    \centering
    \begin{minipage}{0.29\textwidth}
    \includegraphics[width=0.95\textwidth]{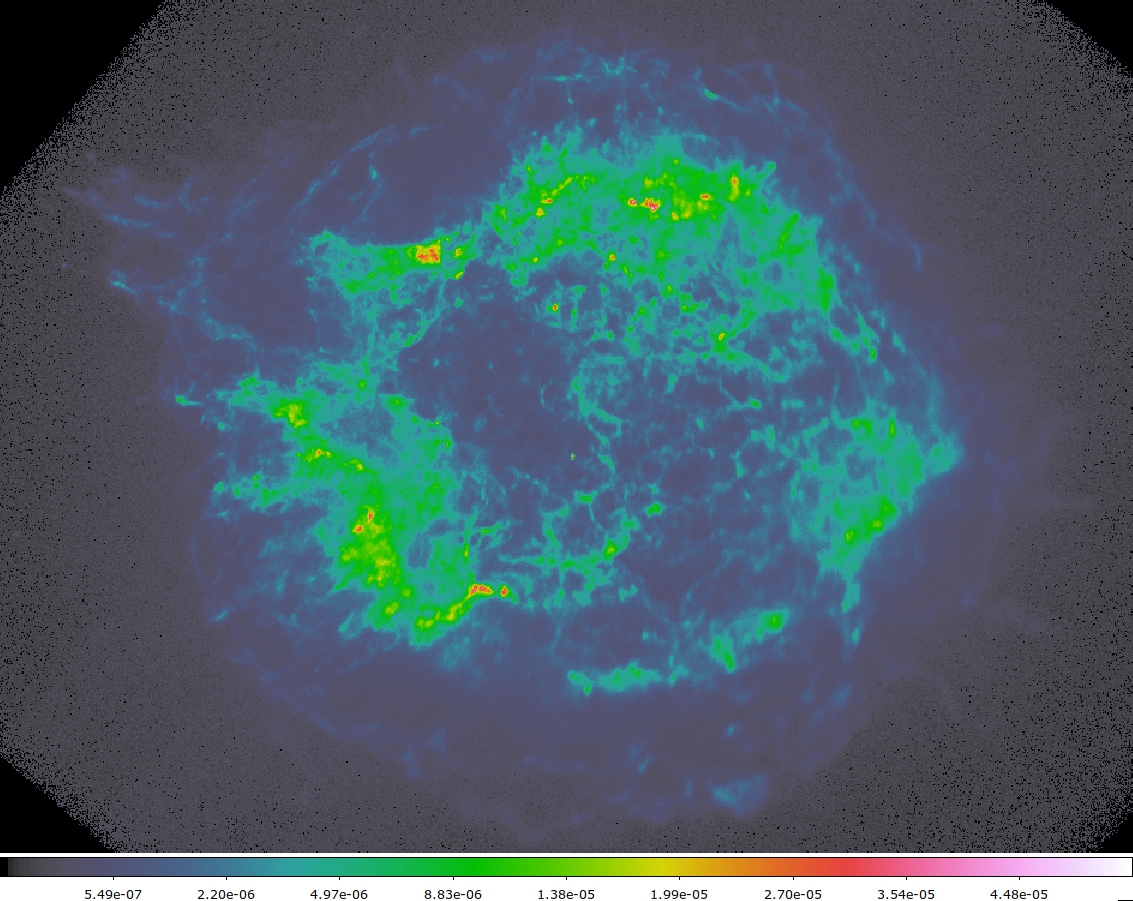}
    \end{minipage}
    \hfill
    \begin{minipage}{0.69\textwidth}
    \includegraphics[width=0.95\textwidth]{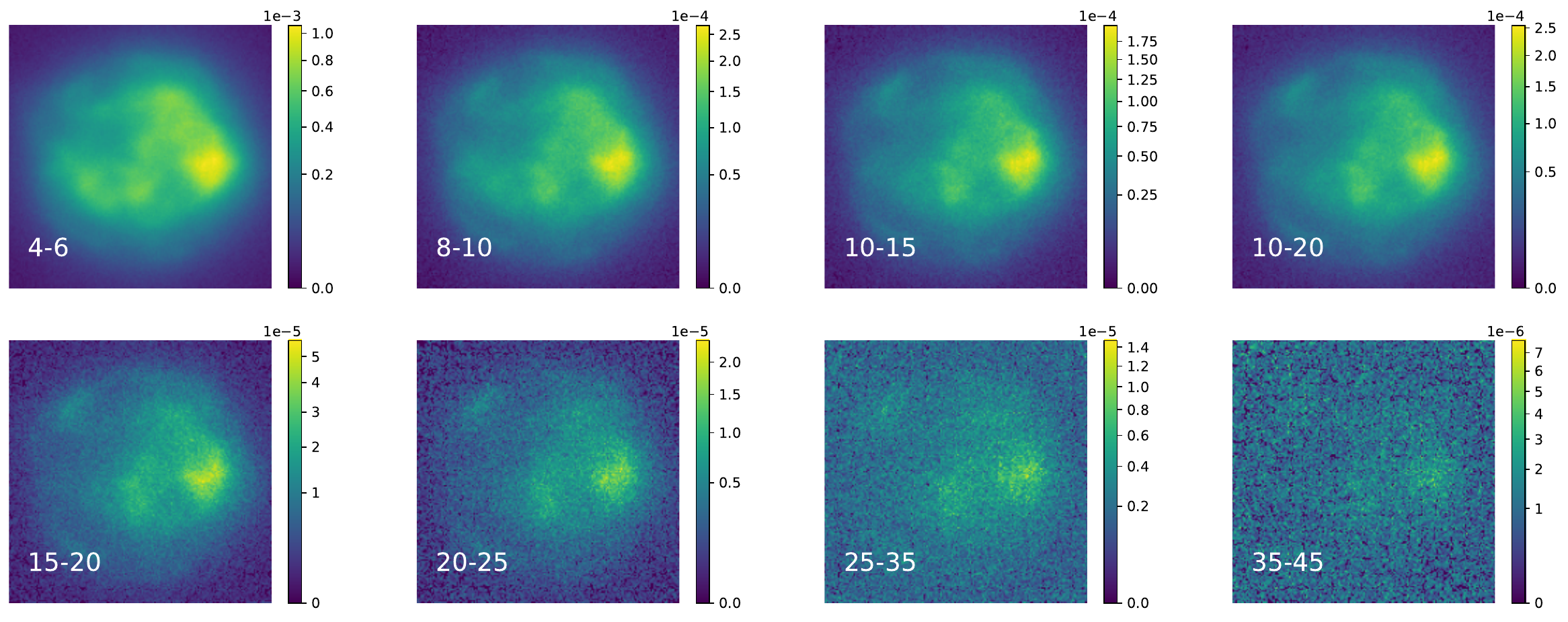}
    \end{minipage}
    \caption{Chandra and NuSTAR exposure/vignetting-corrected images of Cas A. \emph{Leftmost panel.} Chandra/ACIS-S count-rate image in the 0.5-8 keV band with a pixel size of 1'' and a square root scale. \emph{Other panels.} Exposure/vignetting-corrected and mosaicked NuSTAR images of Cas A in various energy bands, reported on the images in units of keV, with a square root scale.}
    \label{fig:acis+nu_images}
\end{figure*}

\begin{figure}
    \centering
    \includegraphics[width=0.95\columnwidth]{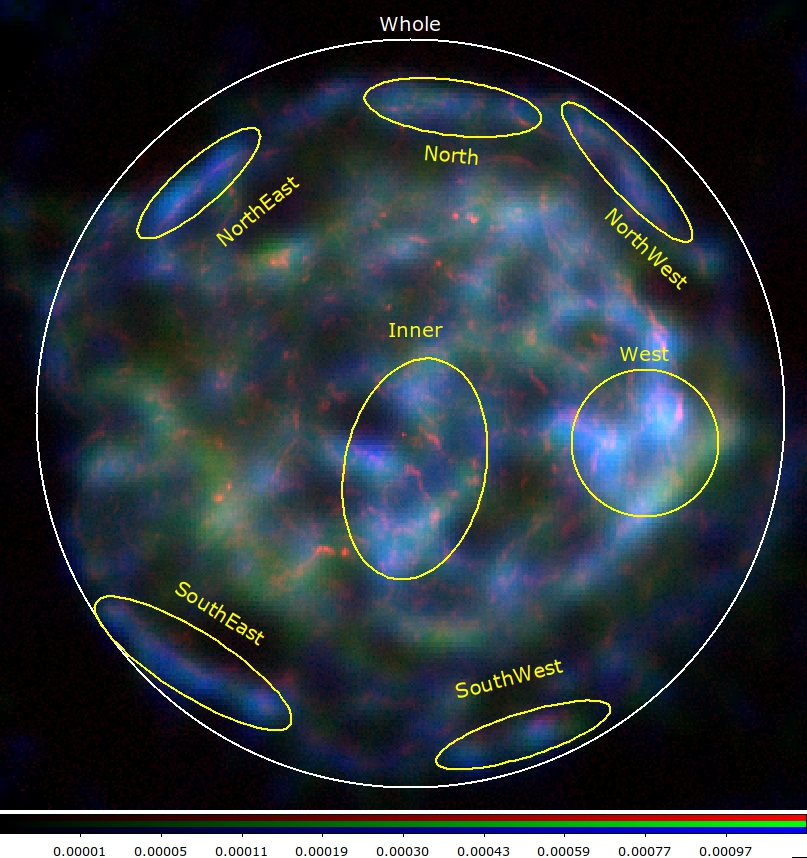}
    \caption{RGB Chandra and NuSTAR image of Cas A with a square root scale. In red, Chandra/ACIS-S count-rate image in the 4-6 keV band with a pixel size of $1''$, in green the 4-6 keV NuSTAR deconvolved count-rate image, in blue the 10-20 keV NuSTAR deconvolved image. The region used for the analysis of the whole remnant is shown in white and the regions selected for the spatially resolved spectral analysis are shown in yellow.}
    \label{fig:region_selection}
\end{figure}

We considered the Chandra/ACIS-S single observation of Cas A with highest exposure time (Obs ID 4638, PI Hwang). We reduced Chandra data with the task \texttt{chandra\_repro} available within CIAO v4.14.0, with CALDB v4.9.7. We produced exposure-corrected images in the 0.5-8 keV and 4-6 keV energy bands with the task \texttt{fluximage}, displayed in the left panel of Fig. \ref{fig:acis+nu_images} and, in red, in Fig. \ref{fig:region_selection}, respectively. 

We used the task \texttt{specextract} to extract the spectra from the Chandra/ACIS-S event list with the proper RMF and ARF files. We selected different background regions in order to cross-check the consistency of our analysis and we verified that the choice of the background did not influence our results. 

\subsection{NuSTAR}
\label{sect:nu}

We analyzed 11 NuSTAR observations of Cas A, listed in Table \ref{tab:obs}, for a total exposure time of more than 2 Ms. The data have been reduced by using the \texttt{nupipeline} task available within NuSTARDAS, version 2.1.2. We set \texttt{SAAMODE=strict} and \texttt{TENTACLE=yes}, in order to be as conservative as possible with the background contamination. We produced exposure and vignetting corrected count-rate images of Cas A in several energy bands, using the same ranges as in \citet{grh15}. We then mosaicked with \texttt{XIMAGE} the images in the same energy band obtained from separate observations and from the two different Focal Plane Modules (FPM) detectors. The resulting mosaicked, exposure and vignetting corrected images are shown, with the corresponding energy band, in Fig. \ref{fig:acis+nu_images}. 

The images shown in the right panel of Fig. \ref{fig:acis+nu_images} were deconvolved through the task \texttt{arestore}, based on the Lucy algorithm (\citealt{luc74}, \citealt{ric72}), 
in order to show that the morphology above 10 keV is consistent with the filamentary structure detected by Chandra.
The comparison between the deconvolved NuSTAR (in green) and the Chandra (in red) images in the 4-6 keV band is shown in Fig. \ref{fig:region_selection}. 
Indeed, the deconvolved NuSTAR count-rate image of Cas A in the 4-6 keV closely matched the same image observed by Chandra, 
proving the reliability 
of our region selection procedure.
The main criterion for selecting the extraction regions was the high surface brightness both in the 4-6 keV and 10-20 keV bands, highlighted (in blue) in Fig. \ref{fig:region_selection}. This is crucial, since  high statistics over a wide energy range is required in order to robustly discriminate between spectra with and without cutoff, especially at energies higher than 10 keV. We highlight that, though the deconvolved NuSTAR image show filamentary structures smaller than the region selected, we could not safely reduce the size of the extraction regions because of the NuSTAR Point Spread Function (PSF). Therefore, we extracted the NuSTAR spectra from the regions shown in Fig. \ref{fig:region_selection} through the task \texttt{nuproducts} following the recipe by \citet{gfh17}. 


\subsection{INTEGRAL}
\label{sect:isgri}

Unlike Chandra and NuSTAR, INTEGRAL/ISGRI does not spatially resolve Cas A and therefore no image analysis is possible. INTEGRAL's pointing strategy involves a dithering on a time scale of a few kiloseconds to reduce coded mask artifacts: each individual pointing is called \emph{science window}. We selected all the 5686 science windows in which the spacecraft was pointing at less than 10 degrees from Cas A. We built a catalog of the 7 detected sources detected at more than 7$\sigma$ in an 20-80 keV image obtained from 10\% of the full sample, randomly selected. Using this catalog, we extracted the average spectrum of the source as seen by the IBIS/ISGRI detector \citep{isgri} using version 11.2 of the Offline Science Analysis \citep{cwb03} served through the Multi-Messenger Online Data Analysis \citep[MMODA,\footnote{\url{ https://www.astro.unige.ch/mmoda/}}][]{integralweb}. The resulting equivalent on-axis exposure is about 10.8 Ms over 17 Ms of observing time. We averaged spectra and responses through the mission: owing to the detector evolution and the relative source faintness, we could reliable use the data from 40 to 100 keV.

\subsection{SWIFT}
\label{sect:bat}

SWIFT/BAT is a coded mask detector that cannot spatially resolve Cas A. We used the spectra available online at \href{https://swift.gsfc.nasa.gov/bat\_survey/bs105mon/spectra/bat\_index_1193.pha}{SWIFT/BAT\_CasA} and the response matrix was downloaded from \href{https://swift.gsfc.nasa.gov/bat\_survey/bs105mon/data/swiftbat\_survey\_full.rsp}{SWIFT/BAT\_RMF}. The total exposure time of the spectra is 180 Ms.


\section{Spectral analysis}
\label{sect:spec_anal}

We performed the spectral analysis with XSPEC (v12.12.1, \citealt{arn96}) in several energy ranges and setups. We analyzed Chandra/ACIS-S data in the 
4-6 keV band; NuSTAR/FPMA,B spectra in the 4-5.5 keV, 9-15 keV and 15-40 keV; INTEGRAL/ISGRI data in the 40-100 keV; SWIFT/BAT observations in the 15-200 keV. All the Chandra and NuSTAR spectra were optimally binned following the procedure described in \citet{kb16} through the FTool task \texttt{ftgrouppha}. The corresponding background spectra were subtracted from Chandra/ACIS-S and NuSTAR/FPMA,B source spectra, whereas the INTEGRAL/ISGRI and SWIFT/BAT spectra were obtained already with proper background subtraction. Our approach in analyzing the spectra is based on testing four different nonthermal scenarios: i) straight power-law without any cutoff; ii)\texttt{srcut} model (\citealt{rk99}); iii) a \texttt{zira} model 
\citep[based on ][see details in Appendix \ref{sect:jitter_xspec}]{za07};
%
iv) a \texttt{jitter} model, that we incorporated 
in XSPEC (see details in Appendix \ref{sect:jitter_xspec}), which represents 
the jitter scenario. 
In order to select the model providing the best description of the data we used three metrics: the standard $\chi^2$ statistic and its reduced version $\chi^2_r=\chi^2/n$ with $n$ the degrees of freedom; the Bayesian Information Criterion (BIC, \citealt{sch78}), defined as BIC$=\chi^2 + m\ln(n)$, $m$ being the number of free parameters in the model; the Akaike Information Criterion \citep{aka74}, defined as AIC$=\chi^2 +2m$.  The model best describing the data is the one with the corresponding lower value of either $\chi^2_r$, AIC or BIC metrics.

As discussed in Sect. \ref{sect:intro}, the overall spectrum of jitter radiation can be divided in two components: the low-energy part, i.e. for $\omega < \omega_{break}$, is similar in shape to the classical synchrotron component; the high energy part, i.e. for $\omega > \omega_{break}$, which can extend up to $\gamma$-ray energies (K13). In the following, we will refer to the low-energy component as \emph{synchrotron component/regime}, to the high-energy component as \emph{jitter component/regime}, while we keep the terms \emph{synchrotron radiation} and \emph{jitter radiation} when referring to the global radiation in the case of uniform and turbulent magnetic field, respectively. It is evident that in order to firmly detect jitter radiation we need to look for the absence of a cutoff in the spectra and a steepening of the photon index between two 
energy bands and this is achievable only by considering wide spectral ranges.

This section is divided in various subsections, corresponding to the different regions of the SNR shown in Fig. \ref{fig:region_selection}.

\subsection{Whole remnant}
\label{sect:whole}

In this subsection we analyze the X-ray spectrum emitted by whole Cas A and extracted from the white region shown in Fig. \ref{fig:region_selection}. In this scenario, we can exploit the information collected by all the telescopes considered in this project. 

\subsubsection{Hard band}
\label{sect:hard_xray}

We first focused on the hard energy band, $> 15$ keV, where the X-ray thermal contribution is negligible and the total measured flux is ascribable to purely nonthermal emission processes. We only considered NuSTAR spectral bins where the background emission is a small fraction ($\lesssim 10\%)$ of the Cas A emission, till an energy of 40 keV.

We performed the analysis either by fitting a single combined NuSTAR spectrum, obtained by processing the original NuSTAR spectra through the \texttt{addascaspec} task, and by simultaneously fitting the NuSTAR spectra extracted from the 11 observations. These two slightly different approaches have pro and cons which balance each other out. Summing the spectra lead to a single global spectrum with much higher statistics, smaller error bars and, therefore, to higher sensitivity to the spectral model adopted. However, since there were 11 separate NuSTAR observations performed in a time-lapse of 2 years, the response matrix and the characteristics of the nonthermal emission might slightly change between the first and the last observations. On the other hand, the simultaneous analysis of the single spectra extracted from each observation, provided very reliable results, at the expense of the sensitivity to the spectral model.

We applied the spectral models introduced above to fit the X-ray synchrotron spectra of Cas A, each coupled to a component modeling the Galactic absorption (\texttt{TBabs} model in XSPEC) and to two gaussians, accounting for the radioactive $^{44}$Ti lines observed in Cas A at 68 keV and 78 keV\citep[e.g.,][]{ghb14}. 
We also included a constant factor to take into account small variation on the extracted spectra due to PSF effects, cross-calibration (see \citealt{mhm15}) and to the temporal offset.

\begin{figure*}[!ht]
    \centering
    \begin{minipage}{0.48\textwidth}
    \includegraphics[width=.95\textwidth]{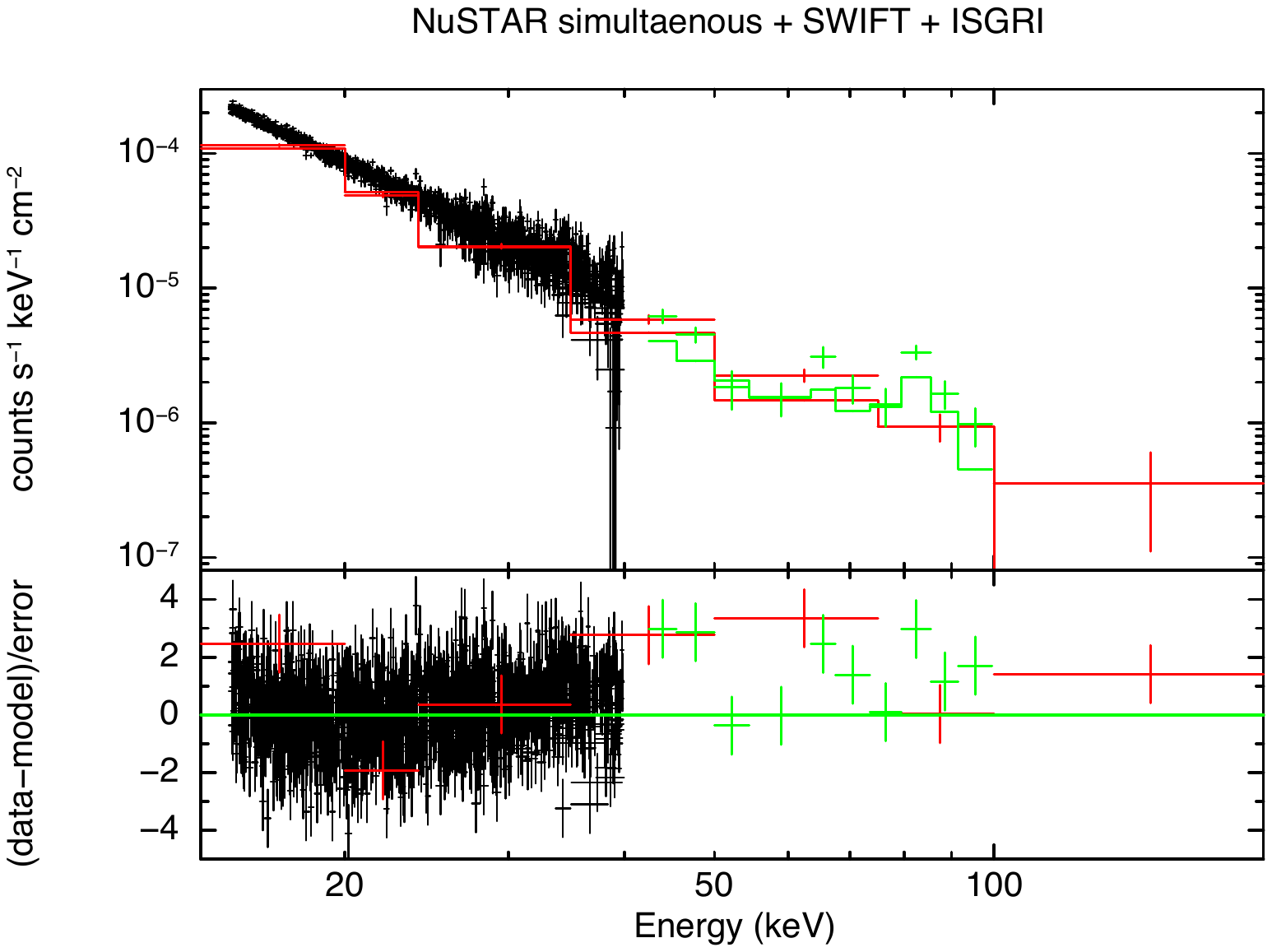}
    \end{minipage}
    \begin{minipage}{0.48\textwidth}
    \includegraphics[width=.95\textwidth]{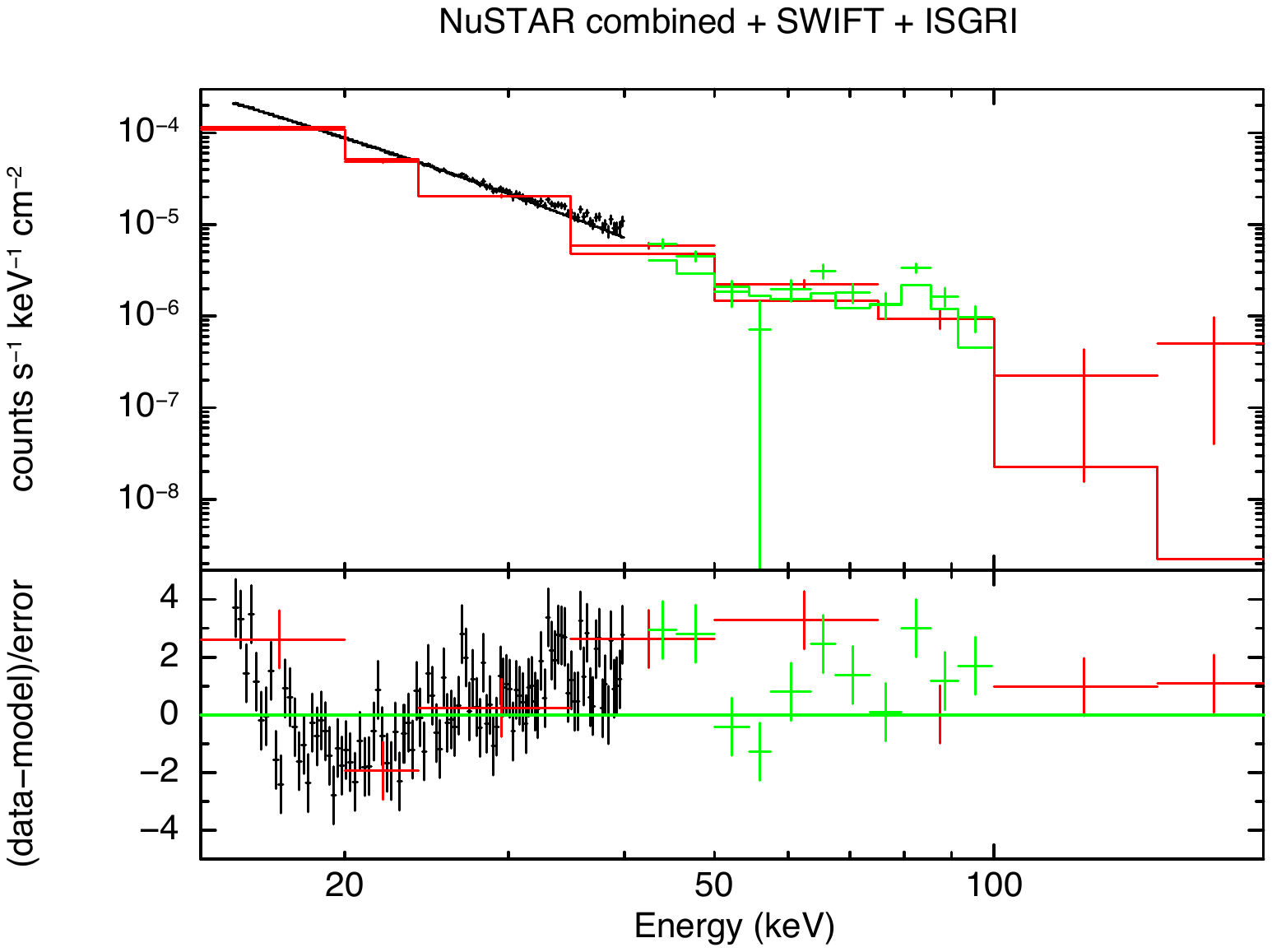} 
    \end{minipage}
    \begin{minipage}{0.48\textwidth}

    \includegraphics[width=.95\textwidth]{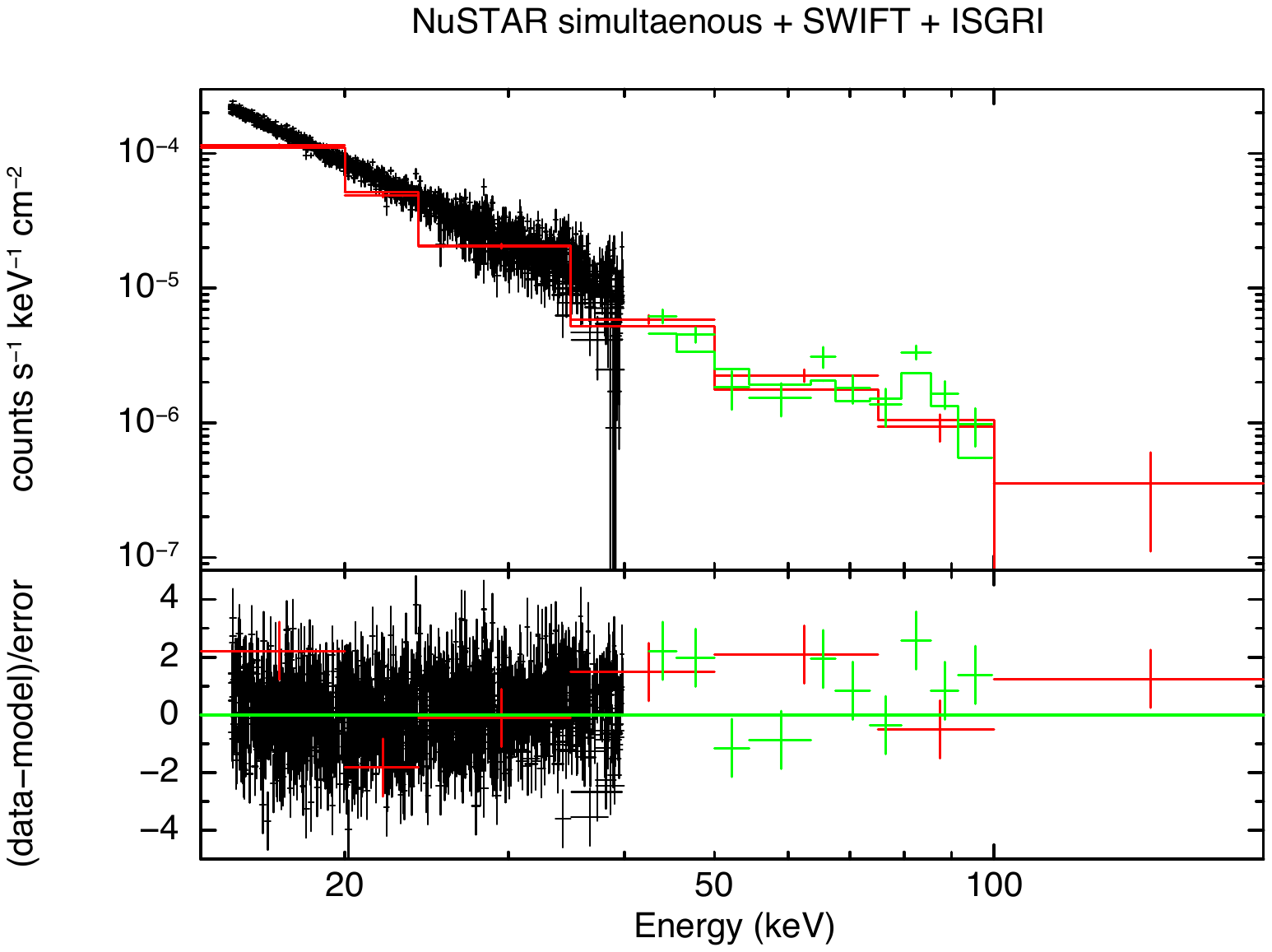}
    
    \end{minipage}
    \begin{minipage}{0.48\textwidth}
    \includegraphics[width=.95\textwidth]{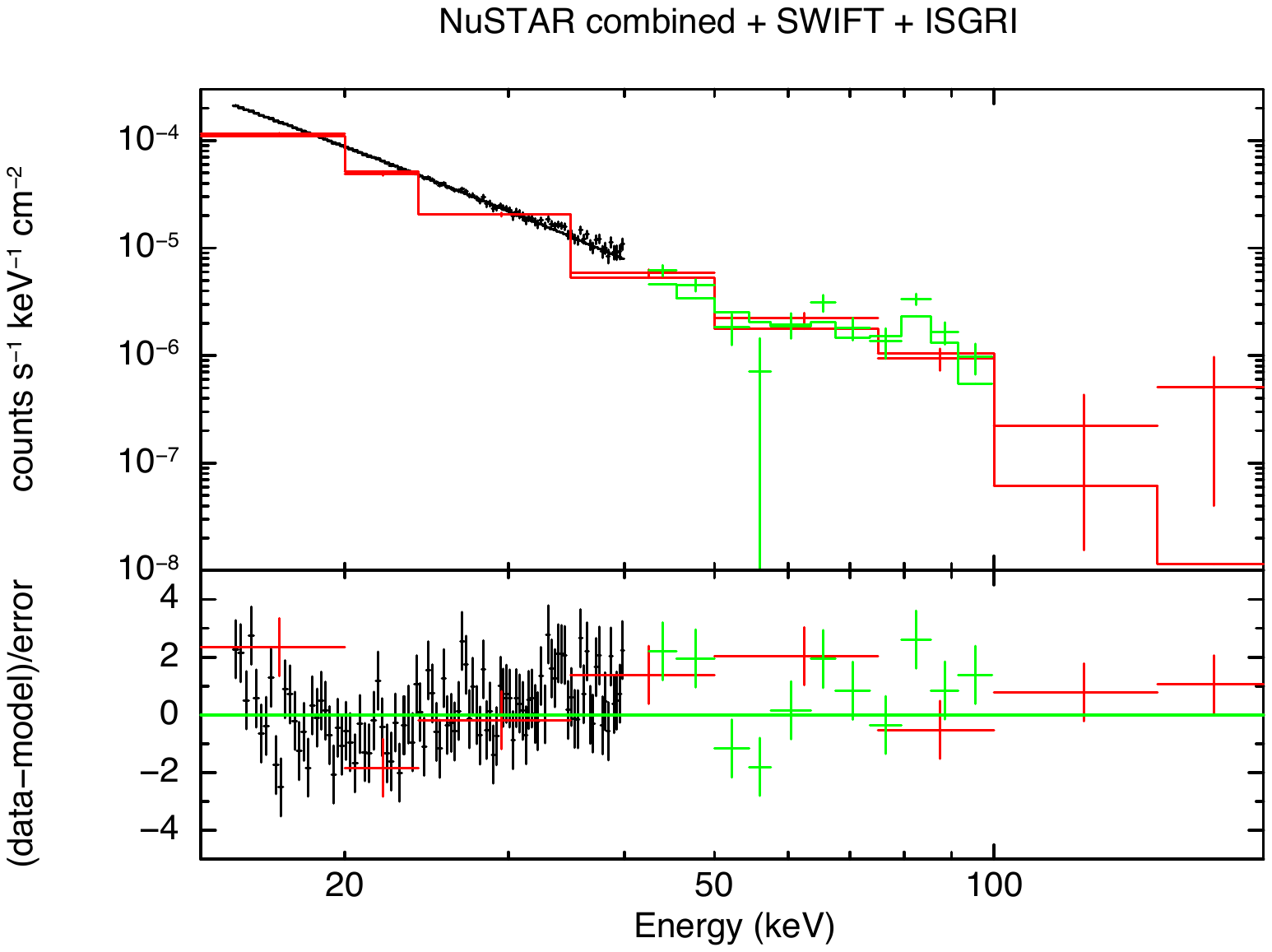}
    \end{minipage}
    \begin{minipage}{0.48\textwidth}
    \includegraphics[width=.95\textwidth]{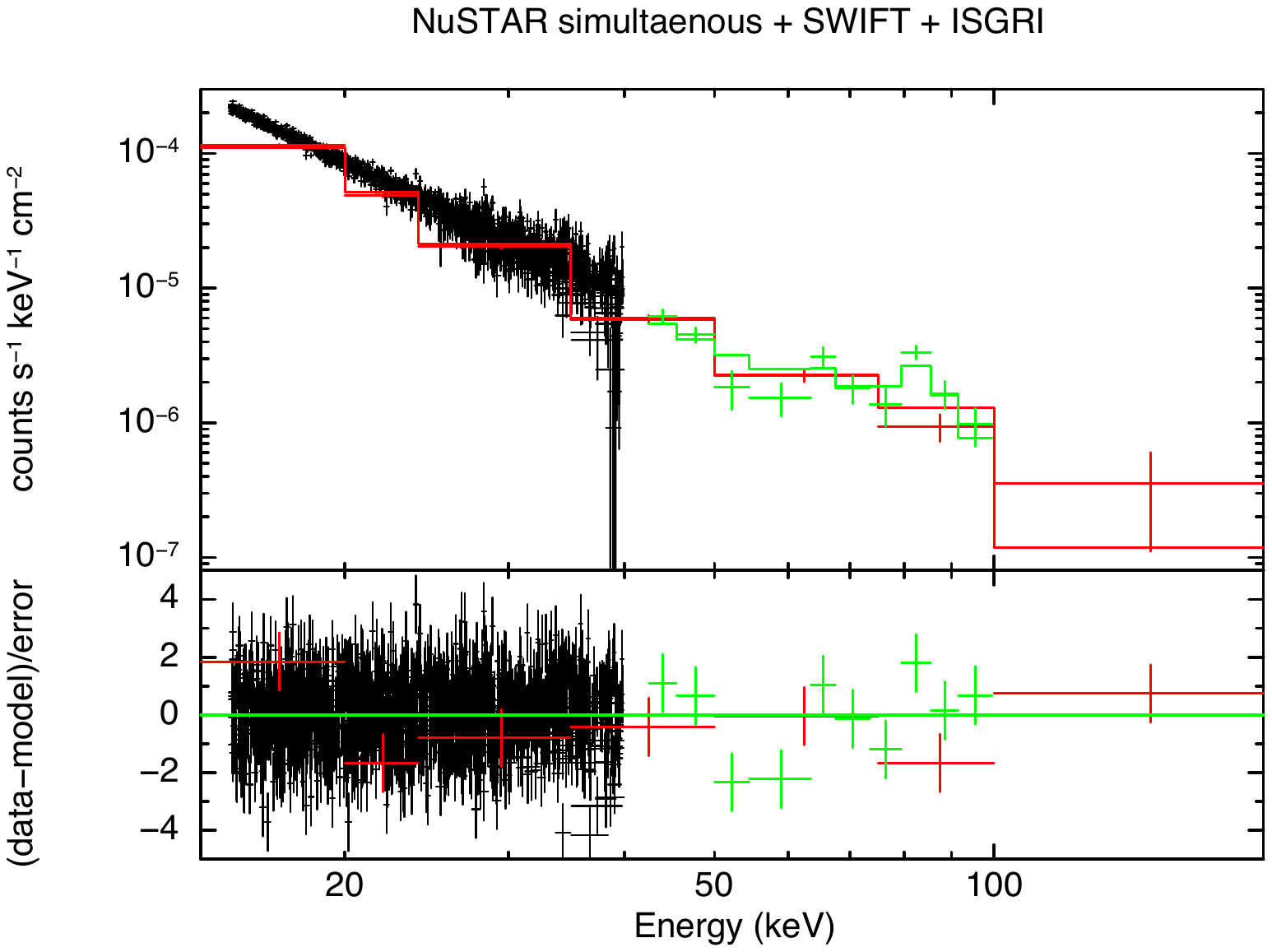}
    \end{minipage}
    \begin{minipage}{0.48\textwidth}
    \includegraphics[width=.95\textwidth]{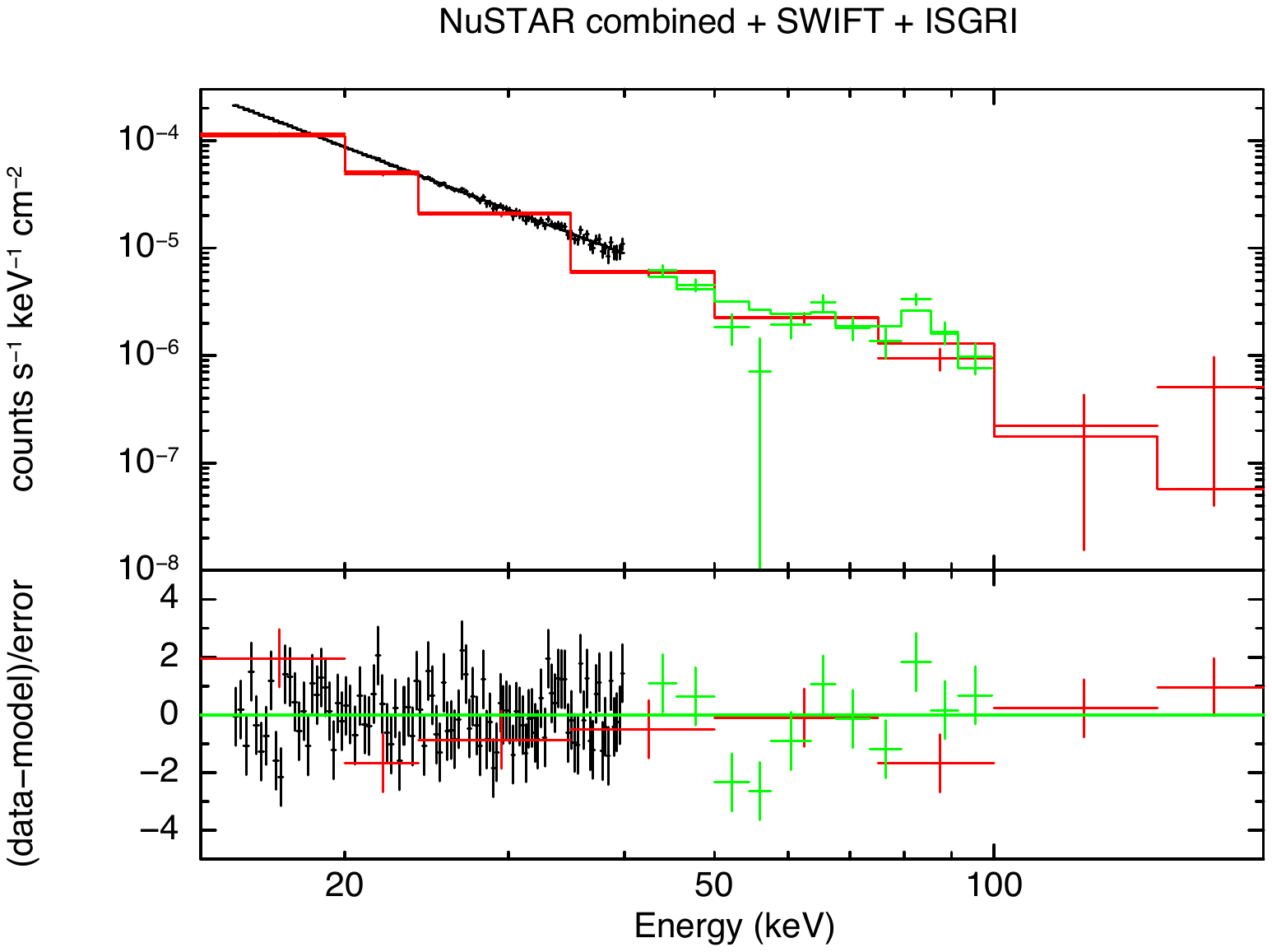}
    \end{minipage}
    \caption{NuSTAR (black), INTEGRAL (green) and SWIFT (red) spectra of Cas A fitted with different models. The normalization of each spectrum is readjusted to the NuSTAR ones through the \texttt{setplot area} command available within XSPEC. Solid line represents the model. \emph{From top to bottom}: spectra are fitted with the \texttt{zira}, \texttt{srcut} and \texttt{pow}/\texttt{jitter} model, respectively (\texttt{jitter} model provides same description as the pow). \emph{Left column.} NuSTAR spectra from all the observations are simultaneously fitted. All the spectra are rebinned for graphical purposes \emph{Right column.} NuSTAR spectra from all the observations are combined.}
    \label{fig:spectra}
\end{figure*}

\begin{table}[!ht] 
    \caption{Best-fit parameters for the 15-200 keV band}
    \centering
    \begin{tabular}{c|c|c|c}
     & Parameter& Sim & Comb\\
    \hline
    & norm$^{a}$  &   0.192$\pm 0.004$ & 0.191 $\pm 0.04$ \\
    \texttt{zira} & E$_{\rm{cut}}$ (keV)  & 1.064$^{+0.018}_{-0.017}$ &1.08$\pm 0.02$ \\
    & $\chi^2$/d.o.f. & 2479/1869 &317/120\\
    & AIC & 2483& 321\\
    & BIC & 2494& 327\\
    \hline
    & norm$^{b}$  & 2180$_{-100}^{+90}$ &2140$\pm 0.80$\\
    & Radio index & \multicolumn{2}{c}{0.77 (fixed)}  \\
    \texttt{srcut} & E$_{\rm{break}}$ (keV) &  1.07 $\pm 0.03$& 1.09$\pm 0.03$\\
    & $\chi^2$/d.o.f.  &  2349/1869 &193/120\\
    & AIC & 2353& 197\\
    & BIC & 2364& 203\\
    \hline
    & norm$^{a}$ &  1.97$\pm 0.08$ & 1.94$\pm 0.08$\\
     & $\Gamma$ & 3.336$^{+0.016}_{-0.015}$& 3.327$\pm 0.014$\\
    \texttt{pow} & $\chi^2$/d.o.f.  & 2270/1869& 129/120 \\
    & AIC & 2274& 133\\
    & BIC & 2285& 139\\
    \end{tabular}
    
    \textbf{Notes.} $^{a}$Units of photons/keV/cm$^2$ at 1 keV. $^{b}$ Flux at 1 GHz in units of Jy. \textit{Simult} and \textit{Comb} mark the results obtained with the simultaneous analysis and with the combination of the NuSTAR spectra, respectively (see text for details). The jitter model is not reported since it is indistinguishable from a simple power-law.
    \label{tab:whole_hard}
\end{table}

We left free to vary the following parameters for each nonthermal model: photon index $\Gamma$ and normalization for the power-law; energy break E$_{\rm{break}}$ and normalization for the \texttt{srcut} model, with the radio spectral index $\alpha$ fixed to 0.77 (\citealt{gre19}); cutoff energy E$_{\rm{cut}}$ and normalization for the \texttt{zira} model. For the jitter model, we left free to vary the photon indices $\Gamma_{1}$ and $\Gamma_{2}$, the normalization, the cutoff shape parameter $\beta$ and the energy break E$_{\rm{break}}$ whereas the ratio between the normalization of the two components was kept fixed to 1.
Values of reduced $\chi^2$ and best-fit parameters for each adopted model are shown in Table \ref{tab:whole_hard}. Fig. \ref{fig:spectra} show the NuSTAR, INTEGRAL and SWIFT spectra fitted with the different models and the corresponding residuals.
The power-law model provided the best description of the spectra at a very significant ( $>5\sigma$) confidence level, as can be clearly seen by the residuals in Fig. \ref{fig:spectra} relative to the \texttt{zira} and \texttt{cutoff} models fitted on the combined spectra. 
The simultaneous analysis provided best-fit parameters perfectly consistent with those obtained with the combined spectra, guaranteeing the reliability of our results. The absence of any cutoff in such a wide (15-100 keV) energy range is the first hint of jitter radiation being a potentially important mechanism for the nonthermal X-ray emission from Cas A . If the turbulence length scale $\lambda$ is smaller than the gyroradius but larger than the photon formation length $\lambda_{pf}$, the slope of the spectrum is directly linked to the exponential of the magnetic-field distribution $\sigma_B$ (see K13) rather than to the turbulence spectrum. Given the purpose of this work, we will focus on the jitter scenario, i.e. assuming that $\lambda < \lambda_{pf}$. 
To investigate the jitter scenario, we fitted the same spectra with the \texttt{jitter} model, finding that the synchrotron component is unconstrained and that the jitter component has same characteristics as the simple power law scenario, i.e. it is not possible to distinguish between a single power-law and a jitter model. By interpreting the 15-100 keV emission as the \textit{jitter component}, the synchrotron component must then be detectable at lower energies, outside this range\footnote{The 100-200 keV band is covered only by SWIFT and because of the wide error bars only upper limits on the flux above 100 keV could be estimated (see Appendix \ref{sect:swift_upperlimit}).}. Naturally, this is true not only for this hard energy range but for a generic X-ray spectrum. Given that the jitter component necessarily comes with a low-energy counterpart (the synchrotron component), the absence of a cutoff in the spectra simply reflects that the synchrotron regime is dominant at energies/frequencies lower than the ones considered. In this framework, the best-fit photon index of $\Gamma \sim 3.3$ implies a turbulence spectrum with an index $\nu_B= 2.3$, higher than typical values for Kolmogorov $\nukol=5/3$ and Kraichnan $\nukra=3/2$ turbulence. It is worth noticing that if $\lambda_{pf} < \lambda < R_{L,j}$, then $\Gamma \sim 3.3$ would imply $\sigma_B \sim 4.3$, from Eq. A5 in K13. Going into details of implications for this scenario is out of the goals of the paper and we leave it for a future study. We will discuss possible explanations for this discrepancy in the jitter scenario in Sect. \ref{sect:turbulence_shape}.

\subsubsection{Soft band}
\label{sect:soft_xray}

In Sect. \ref{sect:hard_xray} we focused on the hard ($\gtrsim 15$ keV) band of the spectra, finding hints for only one of the two components expected in the jitter radiation. If this component is actually the jitter component, we expect the synchrotron one to contribute at lower energies. The typical energy range in which soft X-ray nonthermal emission is studied in SNRs is between 4 and 6 keV, where there are no bright emission lines\footnote{Due to the poor spectral resolution of NuSTAR, some tail of the Fe K emission line contaminated the region between 5.5 and 6 keV. Therefore, we performed the NuSTAR spectral analysis between 4 and 5.5 keV.}.

We adopted the same methodology used for the hard energy band, i.e. we analyzed the Chandra and NuSTAR spectra in the 4-6 keV range with each of the different nonthermal emission models.  We included an absorbing component (\texttt{TBabs} model in XSPEC), though its effect is 
minor at 4 keV. We kept the N$_H$ value fixed to $0.9 \times 10^{22}$, a value found from performing a fit of the Chandra 0.5-8 keV spectra. Given the narrow energy band, we did not consider the jitter model for the 4-6 keV range. The best-fit values with the corresponding errors bars (at 90\% confidence level) of $E_{\rm{break}}$, $\Gamma$ and $E_{\rm{cut}}$ are displayed in the rightmost set of points in Fig. \ref{fig:plot_parnonterm}. 

The best-fit values of $E_{\rm{break}}$, $\Gamma$ and $E_{\rm{cut}}$ show that the 4-6 keV NuSTAR and Chandra spectra are flatter than above 15 keV, where the nonthermal models adopted require higher values of $\Gamma$, $\ebreak$ and $\ecut$. While the steepening $\Delta\Gamma \approx 0.4$ is compatible with what we expect in the jitter framework, it is surprising to observe an increase of the break/cutoff energy with increasing energy range, challenging the picture predicting a unique cutoff for the spectrum of the Whole Cas A, and in agreement with the results obtained in the 15-100 keV band. 

\subsection{Spatially resolved spectral analysis}
\label{sect:internal}

\begin{figure*}[!ht]
\centering
\includegraphics[width=.95\textwidth]{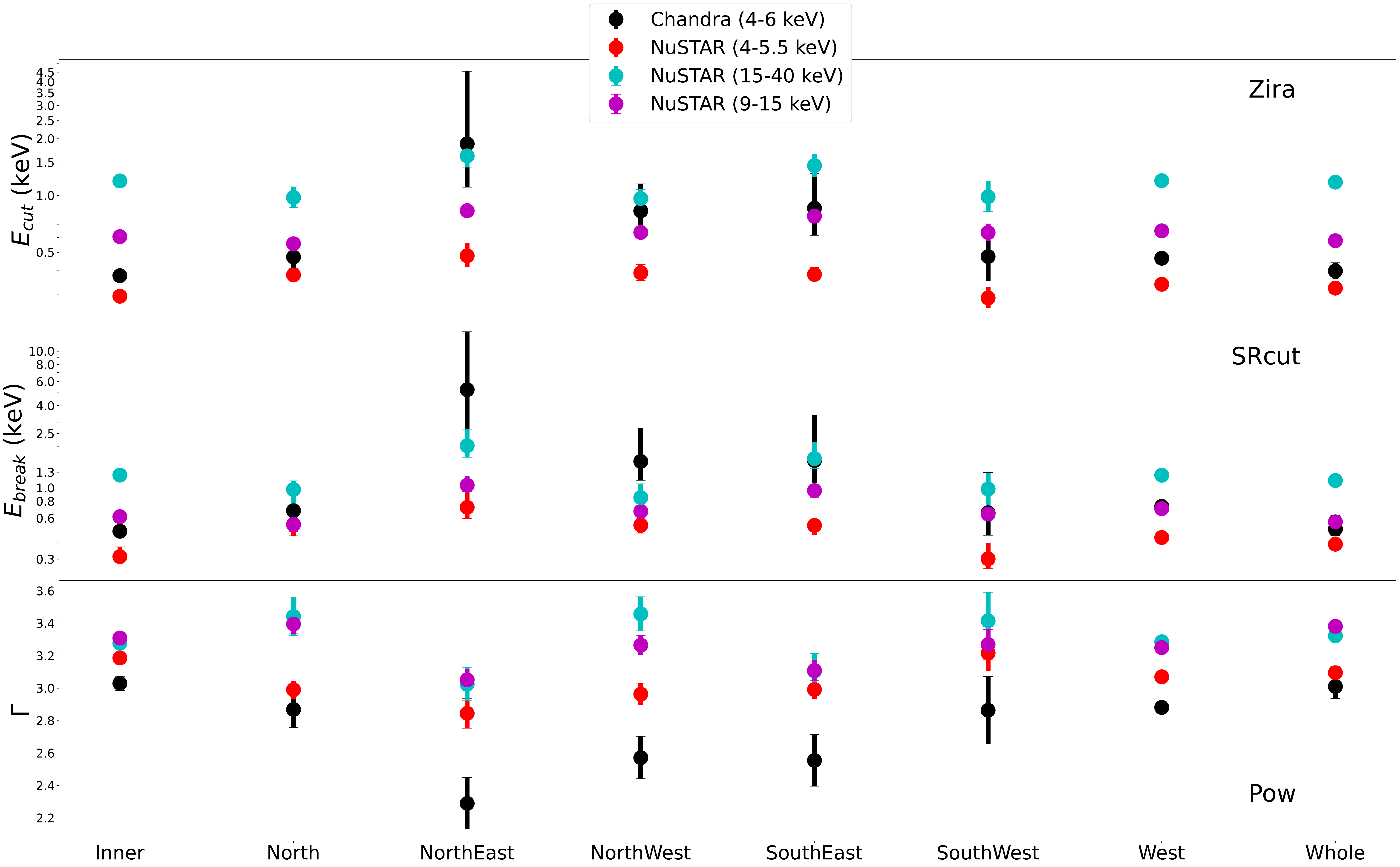}
\caption{Best-fit values of the nonthermal parameters in the \texttt{zira}, \texttt{srcut}, and \texttt{power-law} scenarios, as a function of the region and for different analysis setups. Error bars are estimated at 90\% confidence level. Each color represent the instrument used and the energy range considered: black for Chandra in the 4-6 keV band; red for NuSTAR data in the 4-5.5 keV band; purple for NuSTAR in the 9-15 keV band; cyan for NuSTAR in the 15-40 keV band. }
\label{fig:plot_parnonterm}
\end{figure*}

\begin{figure}[!ht]
\centering
\includegraphics[width=.95\columnwidth]{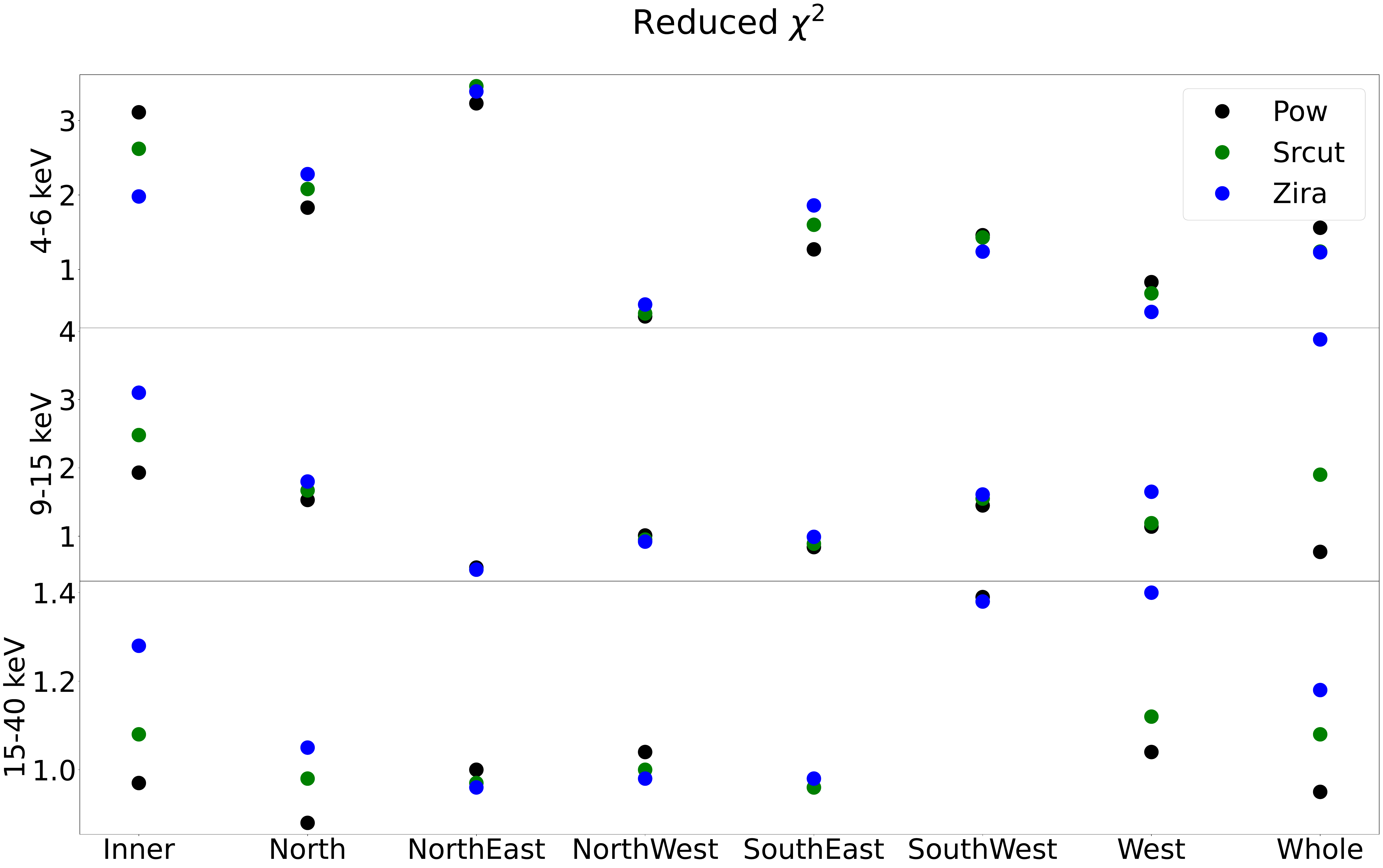}
\caption{Values of $\chi^2_{r}$ for fits performed with different nonthermal models in the 4-6 keV, 9-15 keV and 15-40 energy bands of the NuSTAR spectra. Black, green and blue represent models with a power-law, \texttt{srcut} and \texttt{zira} component, respectively.}
\label{fig:plot_chisq}    
\end{figure}

In Sect. \ref{sect:whole} we showed the results 
of the analysis of the global X-ray spectrum of Cas A,
providing us with information about the overall remnant. In order to investigate the spatial variation of the parameters characterizing the nonthermal emission, we extracted Chandra and NuSTAR spectra\footnote{SWIFT and INTEGRAL are not able to spatially resolve Cas A and cannot be used for this analysis} from the other seven regions shown in yellow in Fig. \ref{fig:region_selection}. The main criterion for the region selection was the high surface brightness both in the 4-6 keV and 10-20 keV that lead to us the identification of five regions close to the shock front: NorthWest (NW), North (N), NorthEast (NE), SouthEast (SE), SouthWest (SW). Moreover, we also considered two additional regions, Inner (I) and West (W): the former being characterized by particularly bright clumps at energies higher than 15 keV, the latter corresponding to the 
location where the emission of nonthermal X-ray radiation indicates fast electron acceleration at the reverse shock
(see e.g. \citealt{owj22,vpc22}). 

We adopted the same procedure used for the whole remnant, by analyzing the Chandra and NuSTAR spectra in the 4-6 keV band and the NuSTAR spectra in the 9-15 and 15-40 keV bands. We here focus only on the spectra combined, in order to increase statistics particularly at energies above 20 keV. 
The best-fit parameters are shown in Fig. \ref{fig:plot_parnonterm} and the corresponding reduced $\chi^2$ values are shown in Fig. \ref{fig:plot_chisq}.

The best-fit values of the nonthermal parameters (with the corresponding 90\% confidence error bars) 
provide the following relevant information:
i) the cut-off energy measured with the \texttt{zira} model systematically increases with increasing energy; ii) the break energy measured with the \texttt{srcut} model shows the same trend but less significantly than the \texttt{zira} scenario, with the NW region marginally consistent at a 90$\%$ confidence level with a fixed break energy; iii) most of the regions are 
characterized by a steepening of the photon index; iv) independently of the spectral model adopted, the Chandra data were systematically requiring flatter spectra than the NuSTAR spectra 
in the 4-6 keV energy band.

The average best-fit value of \ebreak $\,$ changed from 0.5 keV in the 4-6 keV energy band up to 1.5 keV at energies higher than 9 keV.
A similar argument is valid also for \ecut, with an even higher discrepancy since the spectrum in the \texttt{zira} model falls off quicker than in the \texttt{srcut} model. The fit performed by adopting the simple power-law model provided more accurate information on the steepening of the spectra across the SNR. Half of the regions (SE, SW, NE, I) showed spectra with a constant photon index from 4 up to 40 keV, whereas the other half (NW, N, W and the already discussed Whole remnant) presented a clear steepening of $\Delta\Gamma \sim 0.4$. 

We also measured a significant discrepancy between the best-fit values obtained from the Chandra and the NuSTAR spectra in the 4-6 keV energy band, with NuSTAR predicting steeper slope. We checked that this issue was not due to errors in the spectral extraction procedure (see Appendix \ref{app:cross_calibration}) and we concluded that this is probably an effect of cross-calibration between the two telescopes. Given this inconsistency, we did not consider the Chandra spectra in the broadband 4-40 keV analysis presented in the following and we opted for a more conservative approach, based on the analysis of only the NuSTAR spectra.

Figure \ref{fig:plot_chisq} shows the reduced $\chi^2$ for each of the model adopted as a function of the regions. Since all the models adopted have the same number of free parameters, a direct comparison of the $chi^2$ values allowed us to identify the more appropriate model for each of the spectra. We found no significant preference for spectral models with or without cutoff in the 4-6 keV energy band. In regions N, NE, NW, SW and W different nonthermal scenarios provided the same description already at a 1$\sigma$ confidence level, with a $\Delta\chi^2 \leq 1$ among different models. Region SE was marginally better described by a straight power law, at a confidence level lower than $90\%$ \eg{($\Delta\chi^2 < 2.71$)} and regions I and Whole were described better by a cutoff at the $90\%$ confidence level but not at 3$\sigma$. Overall, these results indicated that it was not possible to robustly state whether the 4-6 keV emission was characterized by a cutoff or a straight power-law. An additional source of uncertainty comes from the contribution of thermal continuum emission, 
which is neglected in our analysis and might contribute to the total flux, though this relative contribution could be as low as 5\% as reported by \citet{vpf22} and it is in any case at least a factor 3 lower than the nonthermal one (see Appendix \ref{sect:nonthermalratio}). 

Similar points were also valid for the 9-15 keV and 15-40 keV bands. Regions N, NE, NW, SE, SW were equally described by any of the spectral models adopted in both the energy bands. A fit with a power-law provided a slightly better $\chi^2$ value in region W though not significant at more than 90$\%$ confidence. Only region I and Whole showed a significant preference for the power-law model both in the 9-15 and 15-40 keV energy bands, at odds with the 4-6 keV range. Overall, the results obtained from 15-40 keV band suggested an absence of cutoff in the spectra for region I, N, W, Whole, whereas the other regions did not allow to discriminate between the three different adopted models. 

The results of the spectral analysis presented so far could be summarized in two main points: i) The $\chi^2$ values did not indicate a significant preference for either cutoff or non-cutoff models. This is most likely due to the narrow energy bands considered: the spectral differences between a cutoff and a non-cutoff models are enhanced when looking at the evolution of the shape in wide energy bands. ii) The best-fit values of the break/cutoff energies increased with energy, pointing against the presence of cutoff in the spectra. iii) Best-fit values of photon indices indicated a steepening in at least half of the regions considered. In order to further investigate this point, we analyzed the curvature of the continuum between the soft (4-6 keV) and hard (15-40 keV) energy ranges (see Fig. \ref{sect:4-55+9-40}). 
 
\subsection{Curvature of the continuum}
\label{sect:4-55+9-40}
\begin{figure*}

    \centering
    \includegraphics[width=.9\columnwidth]{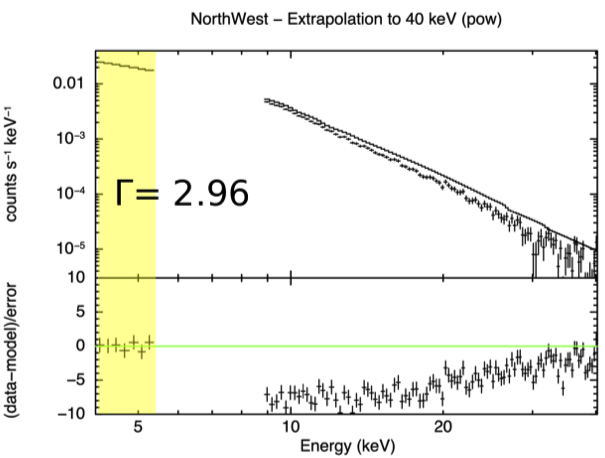}
    \hfill
    \includegraphics[width=.9\columnwidth]{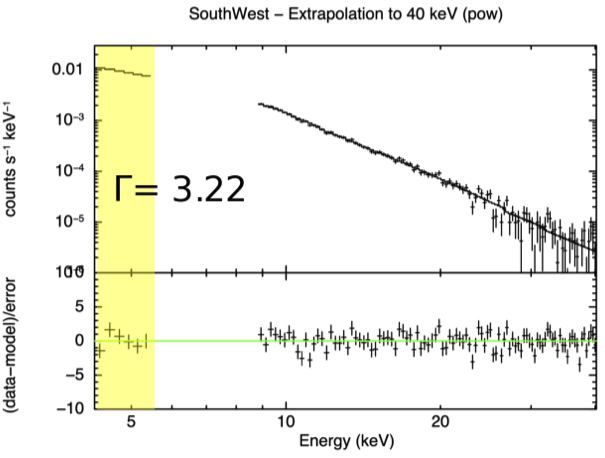}
    
    \includegraphics[width=.9\columnwidth]{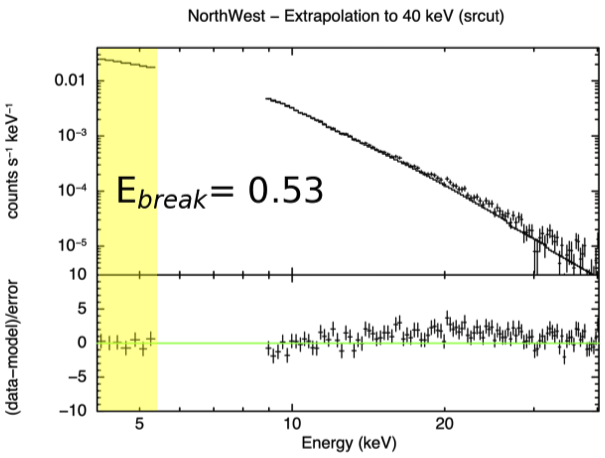}
    \hfill
    \includegraphics[width=.9\columnwidth]{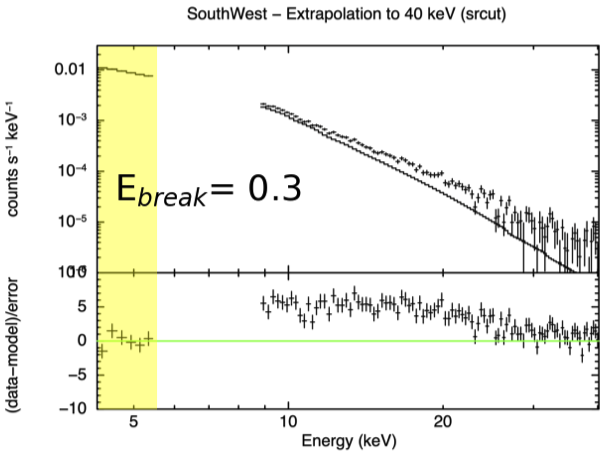}

    \includegraphics[width=.9\columnwidth]{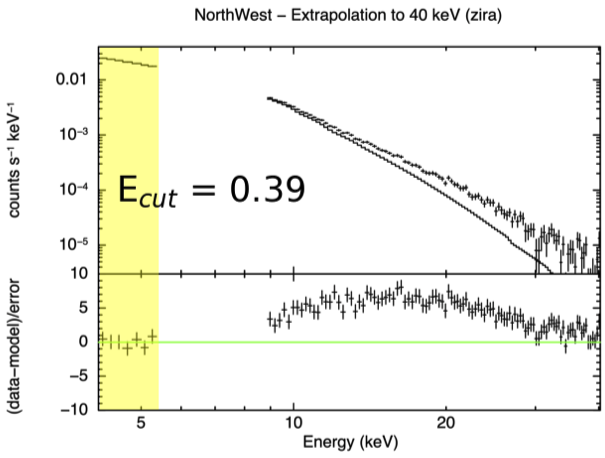}
    \hfill
    \includegraphics[width=.9\columnwidth]{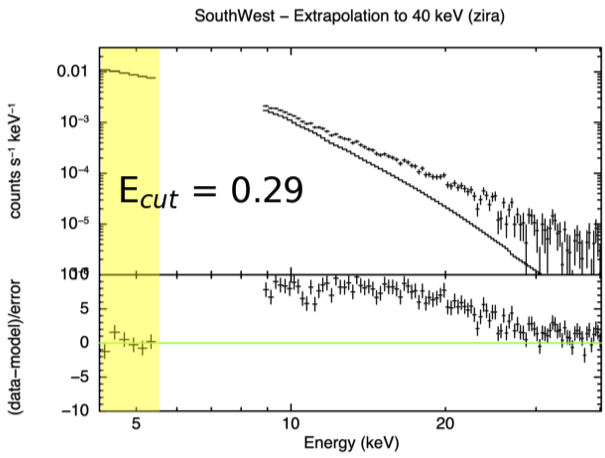}

    \caption{NuSTAR spectra in the 4-5.5 keV and 9-40 keV compared with the best-fit model (solid line) obtained by fitting only the 4-5.5 keV band for regions NW (on the left) and SW (on the right). The yellow area marks the spectral region in which the fit is performed. Each panel shows the 4-6 keV best-fit value of the nonthermal parameter considered: $\Gamma$, $E_{\rm{break}}$ (in keV units) and 
 $E_{\rm{cut}}$ (in keV units) for \texttt{pow}, \texttt{srcut} and \texttt{zira} models, respectively.}
    \label{fig:fixebreak_singleregion}
\end{figure*}

We investigated the change of the curvature of the NuSTAR spectra by extrapolating the best-fit nonthermal component measured from the 4-6 keV range in the 9-40 keV band: if the broadband nonthermal emission was characterized by a cutoff, then the best-fit model in the 4-6 keV band should well describe the data points in the 9-40 keV band as well. In Fig. \ref{fig:fixebreak_singleregion} we show the results of this comparison for regions NW and SW, that were characterized by the two most extreme behaviors, with the best-fit values of the nonthermal parameters obtained from the 4-6 keV band analysis. The same plot for all the regions is available in
Appendix \ref{app:two-bands}.

The adoption of the \texttt{zira} model for the 4-6 keV energy range leads to a heavy underprediction of the higher energy data points, indicating that this kind of cutoff was not suited to describe the broad nonthermal spectrum. The fact that the power-law in the 4-6 keV band was systematically flatter than in the 9-40 keV was already clear in Fig. \ref{fig:plot_parnonterm}. The only regions that did not show a clear steepening from the soft to hard X-rays were NW and SW (see Fig. \ref{fig:fixebreak_singleregion}). Region SW was characterized by the highest $\Gamma \sim 3.2$ and showed the highest underprediction of the data 9-40 keV data points when considering cutoff models, suggesting that in this region the jitter component might be detectable already in the 4-6 keV band, while in other regions it might become so only at higher energies. The spectrum extracted from NW presented an opposite behavior, being well described by the \texttt{srcut} model in the whole 4-40 keV range (see Fig. \ref{fig:fixebreak_singleregion}), indicating that in this region jitter radiation is not at work or that its synchrotron component is dominating the spectrum.   

We also repeated this analysis by fitting the 15-40 keV data and extrapolating the obtained best-fit model down to 3 keV. 
As done in Fig. \ref{fig:fixebreak_singleregion} we here show the plots for SW and NW regions (Fig. \ref{fig:fixebreak_9-40_to_4-6_singleregion}) and we report the results for all the regions in 
Appendix \ref{app:two-bands}. 
 
\begin{figure*}
    \centering  
    \includegraphics[width=.9\columnwidth]{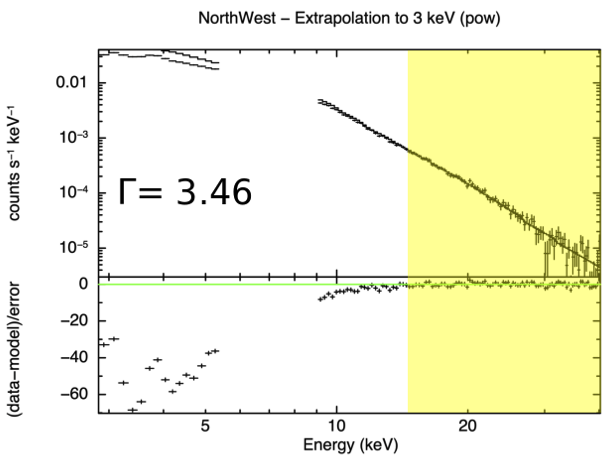}
    \hfill
    \includegraphics[width=.9\columnwidth]{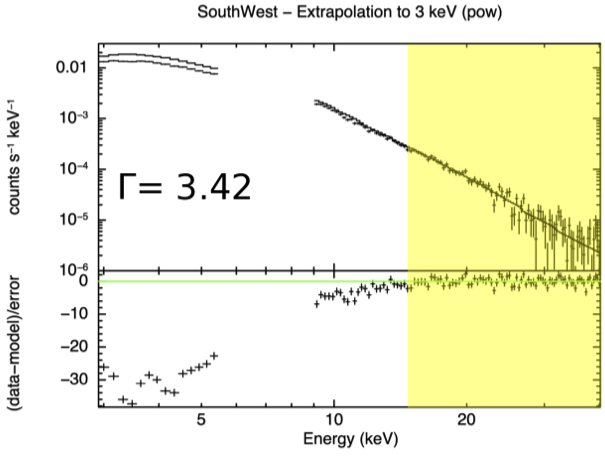}
    
    \includegraphics[width=.9\columnwidth]{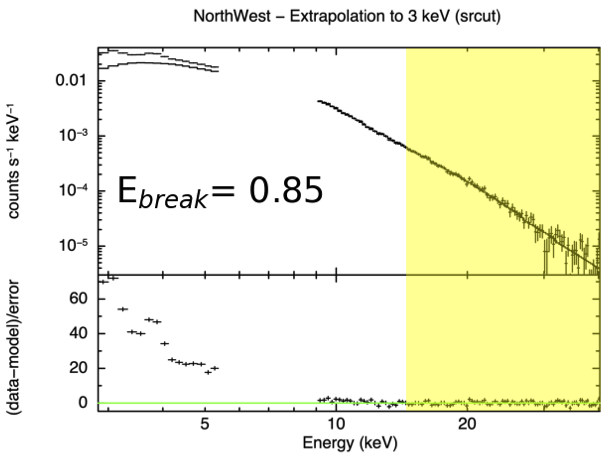}
    \hfill
    \includegraphics[width=.9\columnwidth]{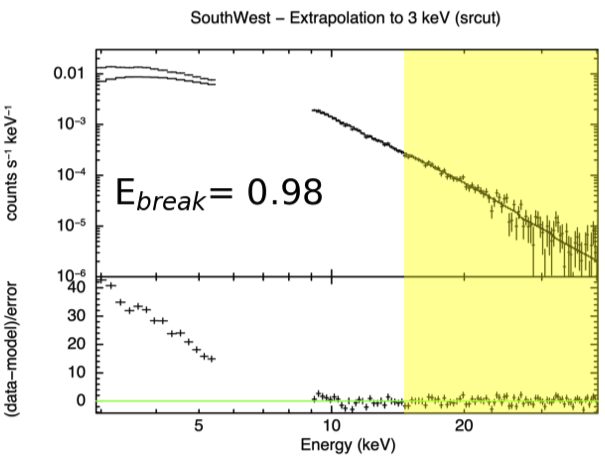}
    
    \includegraphics[width=.9\columnwidth]{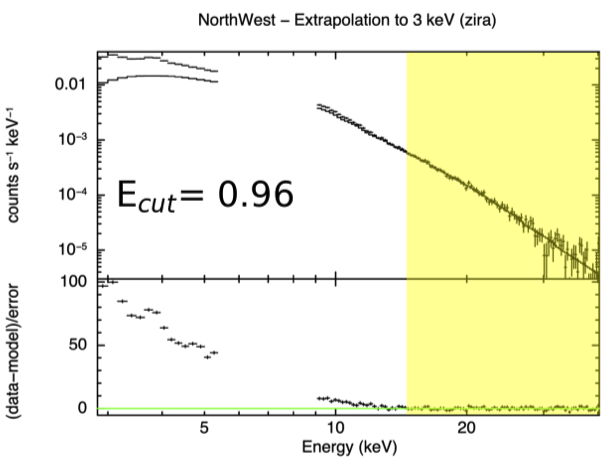}
    \hfill
    \includegraphics[width=.9\columnwidth]{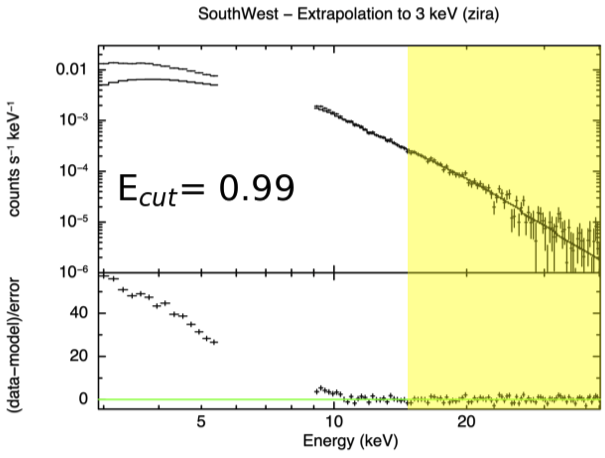}

    \caption{Same as Fig. \ref{fig:fixebreak_singleregion} but here the fit is performed in the 15-40 keV band and extrapolated down to 3 keV. }
    \label{fig:fixebreak_9-40_to_4-6_singleregion}
\end{figure*}

The cutoff models, i.e. \texttt{zira} and \texttt{srcut}, systematically underpredicted the observed data points, with the best-fit value of \ebreak~ and \ecut~ higher than the corresponding values obtained from  the 4-6 keV spectra only (as already observed in Fig. \ref{fig:plot_parnonterm}). The region with the smallest variation in any of these parameter is NW, confirming that its emission is most likely dominated by synchrotron radiation. The extrapolation of the power-law model in region SW lead to a significant overprediction of the data in the 4-6 keV band. 
It is worth mentioning that the results from the 15-40 keV are only indicative and cannot be taken as firm proof of the presence or absence of steepening in the spectra, given that the wide error bars in this range lead to broad uncertainty on the slope of the high-energy tail, as in the case of region SW. In fact, while extrapolating the power-law with $\Gamma=3.4$ resulted in an overprediction of the 4-6 keV band, by taking $\Gamma=3.2$, a value within the 90\% confidence level, the resulting extrapolated power-law very well described the 4-6 keV data points. 

Overall, we found that the spectra face a steepening between the soft and the hard band and that the cutoff models describe the data systematically and significantly worse than the standard power-law. However, the results presented so far did not take into account the shape of the spectrum in the two bands simultaneously. Therefore, 
we also fitted the 
the 4-6 keV and 9-40 keV energy bands 
simultaneously (Sect. \ref{sec:fit_jitter}), excluding the 6-9 keV range that is dominated by the Fe K emission line. 

\subsection{Jitter model}
\label{sec:fit_jitter}
The results presented in Sect. \ref{sect:4-55+9-40} provided a qualitative estimate of how coherent is the continuum shape between the soft, 4-6 keV, and hard, 9-40 keV, energy bands. We fitted the 4-6 plus 9-40 keV energy band for all the regions with our \texttt{jitter} model (see Appendix \ref{sect:jitter_xspec}) and the standard \texttt{srcut}, \texttt{zira} and \texttt{pow} models. We performed the analysis either by leaving the $\beta$ free to vary and by keeping it fixed to 0.5 (as in the \texttt{zira} model), 1 (as in the \texttt{srcut} model) and 2. We found that leaving the $\beta$ free to vary significantly improved the quality of the fit for all the regions but NW, with an average value of roughly 0.5 and systematically lower than 1. The best-fit parameters for all the models with corresponding $\chi^2_{r}$, BIC and AIC are listed in Table \ref{tab:param_freeebreak}. 

\begin{figure*}[!ht]
    \centering
    \begin{minipage}{0.32\textwidth}
    \includegraphics[width=.97\columnwidth]{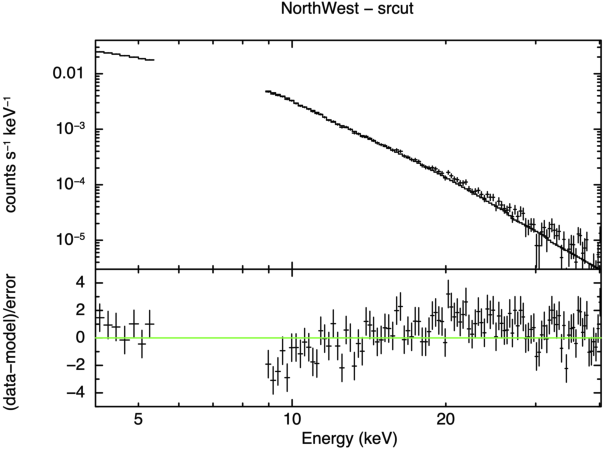}
    \includegraphics[width=.97\columnwidth]{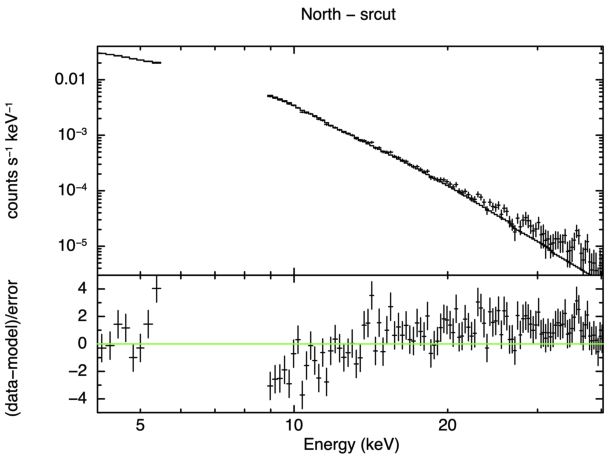}
    \includegraphics[width=.97\columnwidth]{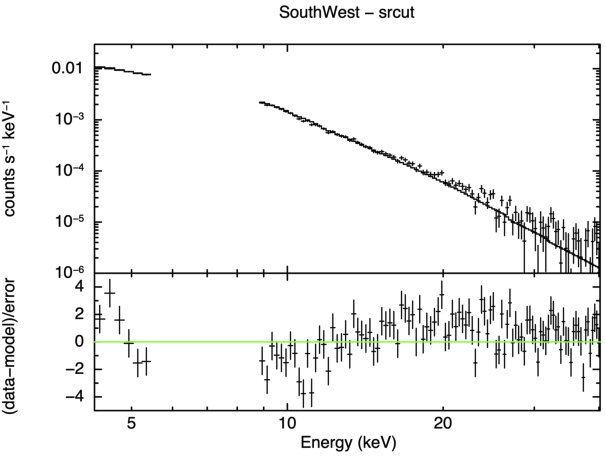}
    \end{minipage}
    \begin{minipage}{0.32\textwidth}
    \includegraphics[width=.97\columnwidth]{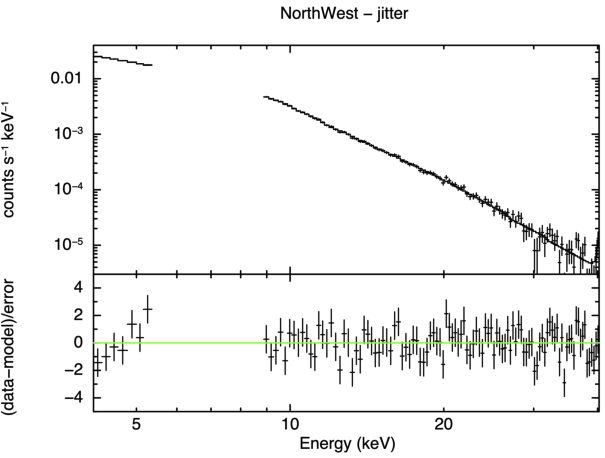}
    \includegraphics[width=.97\columnwidth]{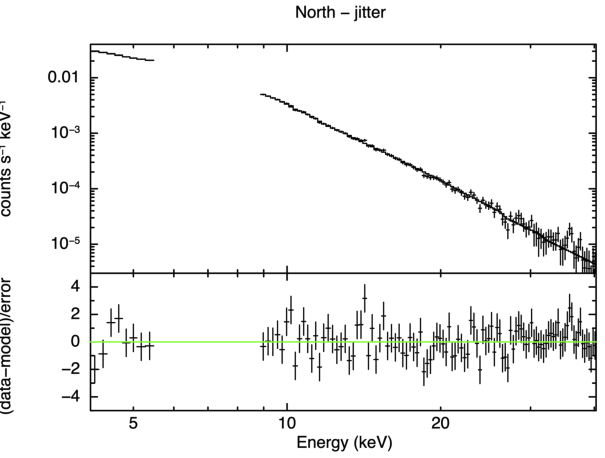}
    \includegraphics[width=.97\columnwidth]{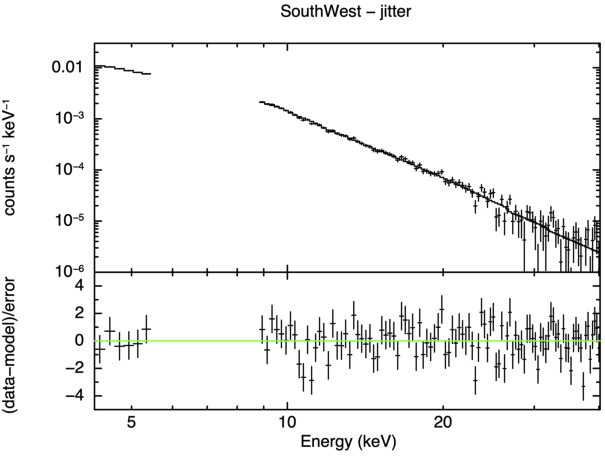}
    \end{minipage}
    \begin{minipage}{0.32\textwidth}
    \includegraphics[width=.97\columnwidth]{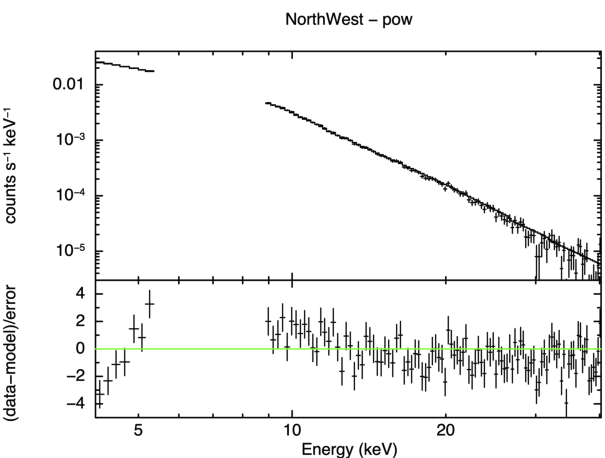}
    \includegraphics[width=.97\columnwidth]{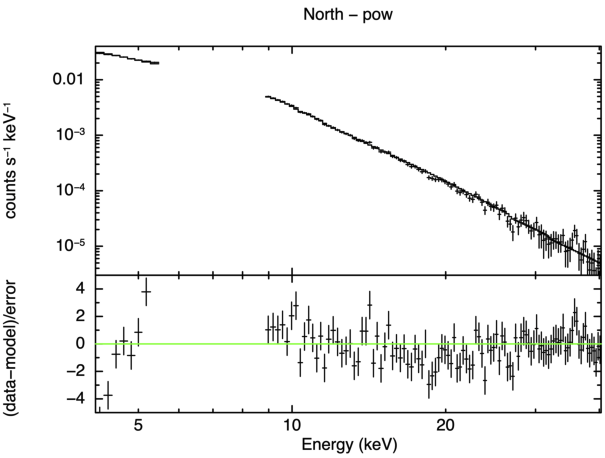}
    \includegraphics[width=.97\columnwidth]{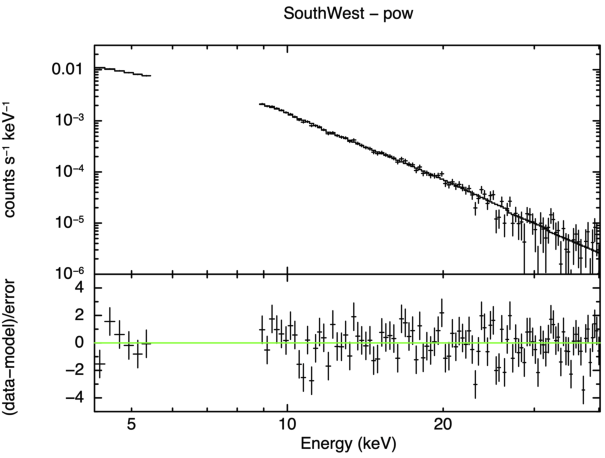}
    \end{minipage}
    \caption{NuSTAR spectra from regions NW (top line), N (central line), SW (bottom line) fitted with \texttt{srcut} (left column), \texttt{jitter} (central column) and \texttt{pow} (right column) models in the 4-5.5 keV and 9-40 keV and corresponding residuals. }
    \label{fig:freebreak_4-9-40_singleregion}
\end{figure*}

\begin{table*}[!ht]
    \centering
    \caption{Best-fit values on the 4-5.5 plus 9-40 keV bands}
    \begin{tabular}{c|c|c|c|c|c|c|c|c|c}
       \multicolumn{2}{c}{ }  &\multicolumn{8}{|c}{Region}   \\
    \hline
    Model &Parameter & Inner & North & NorthEast & NorthWest & SouthEast & SouthWest & West & Whole \\
    \hline
\texttt{TBabs} & N$_H$$^a$ & 1.7 & 1.1 & 1.0 & 1.13 & 1.5 & 1.4 & 2.0 & 0.9 \\
\hline
&\ecut (keV)& 0.447$_{-0.002}^{+0.002}$ & 0.421$_{-0.004}^{+0.004}$ & 0.68$_{-0.01}^{+0.01}$ & 0.51$_{-0.006}^{+0.006}$ & 0.547$_{-0.007}^{+0.007}$ & 0.488$_{-0.009}^{+0.009}$ & 0.497$_{-0.002}^{+0.002}$ & 0.494$_{-0.001}^{+0.001}$ \\ 
 &norm$^b$& 73.2$_{-0.4}^{+0.4}$ & 8.5$_{-0.1}^{+0.1}$ & 3.38$_{-0.05}^{+0.05}$ & 6.03$_{-0.08}^{+0.08}$ & 5.46$_{-0.07}^{+0.07}$ & 2.9$_{-0.06}^{+0.06}$ & 57.0$_{-0.2}^{+0.3}$ & 599.0$_{-2.0}^{+2.0}$ \\ 
\texttt{zira} &$\chi^2$$_r$& 46.5 & 6.5 & 5.1 & 5.25 & 8.25 & 4.42 & 62.32 & 178.75 \\ 
&AIC& 5723.37 & 731.79 & 560.24 & 581.76 & 960.48 & 472.71 & 7918.88 & 25268.8 \\ 
&BIC & 5728.99 & 737.23 & 565.62 & 587.16 & 965.99 & 478.04 & 7924.57 & 25274.99 \\
\hline 
 &\ebreak (keV) & 0.462$_{-0.004}^{+0.004}$ & 0.424$_{-0.007}^{+0.007}$ & 0.9$_{-0.02}^{+0.02}$ & 0.57$_{-0.01}^{+0.01}$ & 0.65$_{-0.01}^{+0.01}$ & 0.53$_{-0.02}^{+0.02}$ & 0.545$_{-0.003}^{+0.003}$ & 0.486$_{-0.002}^{+0.002}$ \\ 
& norm$^c$& 930$_{-10}^{+10}$ & 114$_{-3}^{+3}$ & 28.0$_{-0.8}^{+0.8}$ & 66$_{-2}^{+2}$ & 55$_{-1}^{+1}$ & 33$_{-1}^{+1}$ & 650$_{-5}^{+5}$ & 8050$_{-60}^{+60}$ \\ 
\texttt{srcut} &$\chi^2$$_r$& 17.68 & 2.48 & 2.25 & 1.78 & 3.79 & 2.39 & 19.08 & 62.26 \\ 
&AIC & 2178.41 & 281.75 & 248.94 & 199.77 & 443.95 & 257.58 & 2427.71 & 8710.71 \\ 
&BIC& 2184.03 & 287.19 & 254.32 & 205.17 & 449.46 & 262.91 & 2433.4 & 8716.9 \\ 
 \hline
 &$\Gamma$& 3.295$_{-0.004}^{+0.004}$ & 3.333$_{-0.009}^{+0.009}$ & 2.99$_{-0.01}^{+0.01}$ & 3.186$_{-0.009}^{+0.009}$ & 3.124$_{-0.009}^{+0.009}$ & 3.22$_{-0.01}^{+0.01}$ & 3.219$_{-0.003}^{+0.003}$ & 3.347$_{-0.002}^{+0.002}$ \\ 
& norm$^b$& 198$_{-2}^{+2}$ & 22.8$_{-0.4}^{+0.4}$ & 7.8$_{-0.2}^{+0.2}$ & 15.2$_{-0.3}^{+0.3}$ & 13.3$_{-0.2}^{+0.2}$ & 7.5$_{-0.2}^{+0.2}$ & 150.0$_{-0.8}^{+0.8}$ & 2070$_{-10}^{+10}$ \\ 
\texttt{pow} &$\chi^2$$_r$& 1.5 & 2.28 & 1.05 & 1.83 & 1.11 & 1.38 & 2.75 & 2.14 \\ 
&AIC& 188.13 & 259.9 & 118.1 & 205.77 & 132.39 & 150.58 & 353.74 & 391.93 \\ 
&BIC& 193.75 & 265.34 & 123.48 & 211.17 & 137.9 & 155.91 & 359.43 & 398.12 \\  
 \hline 
& $\Gamma_1$ & 2.87$_{-0.3}^{+0.02}$ & 2.87$_{-0.04}^{+0.02}$ & 2.45$_{-0.05}^{+0.05}$ & 3.0$_{-0.08}^{+0.08}$ & 2.5$_{-0.3}^{+0.3}$ & 2.9$_{-0.07}^{+0.2}$ & 2.71$_{-0.01}^{+0.01}$ & 2.724$_{-0.01}^{+0.006}$ \\ 
& \ebreak (keV) & 11.8$_{-0.3}^{+0.3}$ & 5.2$_{-0.1}^{+0.1}$ & 5.2$_{-0.2}^{+0.3}$ & 40.4$    _{-0.8}^{+0.5}$ & 6$_{-1}^{+4}$ & 4.4$_{-0.4}^{+0.2}$ & 5.36$_{-0.03}^{+0.08}$ & 4.59$_{-0.1}^{+0.04}$ \\ 
&$\beta$& 0.57$_{-0.04}^{+0.04}$ & 0.38$_{-0.02}^{+0.03}$ & 0.6$_{-0.2}^{+0.2}$ & 1.2$_{-0.4}^{+0.7}$ & 0.65$_{-0.06}^{+0.06}$ & 0.3$_{-0.3}^{+0.1}$ & 0.439$_{-0.003}^{+0.02}$ & 0.485$_{-0.009}^{+0.007}$ \\ 
& $\Gamma_2$& 3.27$_{-0.02}^{+0.02}$ & 3.429$_{-0.009}^{+0.02}$ & 3.03$_{-0.02}^{+0.03}$ & 2.6$_{-0.2}^{+0.5}$ & 3.05$_{-0.03}^{+0.04}$ & 3.5$_{-0.3}^{+0.1}$ & 3.277$_{-0.008}^{+0.004}$ & 3.359$_{-0.003}^{+0.004}$ \\ 
\texttt{jitter}& norm$^b$ & 186$_{-10}^{+10}$ & 29$_{-1}^{+2}$ & 8.6$_{-0.4}^{+0.6}$ & 12.2$_{-0.3}^{+0.8}$ & 10.8$_{-0.9}^{+1.0}$ & 12$_{-4}^{+4}$ & 173$_{-2}^{+1}$ & 2130$_{-20}^{+20}$ \\ 
&$\chi^2$$_r$ & 1.19 & 1.06 & 0.95 & 1.03 & 0.93 & 1.38 & 1.14 & 1.04 \\ 
&AIC & 152.99 & 125.79 & 110.94 & 119.85 & 115.13 & 151.74 & 150.94 & 181.61 \\ 
&BIC& 166.93 & 139.25 & 124.26 & 133.21 & 128.77 & 164.91 & 165.04 & 196.99 \\  
\end{tabular}

\textbf{Notes.} $^{a}$ Units of 10$^{22}$ cm$^{-2}$, fixed in the fitting procedure. $^{b}$Units of 10$^{-3}$ photons/keV/cm$^2$/s at 1 keV. $^{c}$ Flux at 1 GHz in units of Jy. 
\label{tab:param_freeebreak}
\end{table*}

The recurrent striking feature evident in Table \ref{tab:param_freeebreak} is that the \texttt{jitter} model systematically provides a better description of the data than any cutoff model, as witnessed not only by the $\chi^2_r$ but also by the BIC and AIC metrics. While in the 4-6 keV and in the 15-40 keV bands analyzed separately we were not able to pinpoint a favored model for the nonthermal emission (see Fig. \ref{fig:plot_chisq}), by simultaneously analysing such a wide spectral ranges we are more sensitive to the emission pattern. Hints for this result were already available in Fig. \ref{fig:plot_parnonterm} where we noticed a (monotonic) increase in the photon index values and break energies with increasing energy ranges. We found values of $\Gamma_1$ slightly, but systematically, lower than those obtained with the analysis of the 4-6 keV band alone (red points in Fig. \ref{fig:plot_parnonterm}). This is not surprising given that the \texttt{jitter} model provides the slope of the power-law corrected by the effect of the cutoff, while a simple power-law applied to a cutoff spectrum leads to an overall softer spectrum. On the other hand, best-fit values for $\Gamma_2$ were perfectly compatible with the $\Gamma$ best-fit values obtained from the analysis of the 15-40 keV spectra with the standard power-law model, since in this regime no curvature is present. Thanks to the simultaneous 4-6 keV and 9-40 keV analysis we were now able to quantify and directly compare different models. We highlight this in Fig. \ref{fig:freebreak_4-9-40_singleregion} where we show a comparison between the \texttt{srcut}, \texttt{zira} and \texttt{jitter} model for the regions NW, N and SW, representing three different regimes. 

All the regions but NW and SW showed indication for the presence of both synchrotron and jitter components of jitter radiation as all the parameters were satisfyingly constrained and the energy breaks lied within the spectral range investigated. Region N is shown in Fig. \ref{fig:freebreak_4-9-40_singleregion} as clearly depicts this most common regime, with $4.5 < \ebreak < 6 $ keV, and an average value of roughly 5 keV. Regions I was characterized by slightly higher, but quite unconstrained \ebreak value. In particular, it was compatible, at a $3\sigma$ confidence level, with 5 keV and could then safely safely included in this class of regions. 
Region NW showed a $\ebreak \sim 40$ keV, at the upper edge of the spectral energy considered indicating that only the synchrotron component was detectable in the spectrum. This result nicely fits the picture discussed in Sect. \ref{sect:4-55+9-40} where NW was the region better described by \texttt{srcut} model. Moreover, this is also the only region better reproduced by such a model rather than by a simple power-law.   
Region SW showed a $\beta$ poorly constrained with a \ebreak~ of $\sim$ 4 keV, in the bottom edge of the spectral range considered, indicating that in this region the synchrotron component of jitter radiation might be non detectable. Again, this feature was already suggested by the plot in Fig. \ref{fig:plot_parnonterm} where no clear steepening was observed between the 4-6 keV and 9-40 keV band for the SW. We also noticed that this was the only region that shows no preference for the \texttt{jitter} model with respect to a classical power-law in terms of $\chi^2$, though the best-fit values of the two jitter photon indices differed significantly from the best-fit found with the single pow-law. We investigated this discrepancy by imposing an energy break lower than 4 keV, representing a pure jitter component scenario. The resulting best-fit model showed a $\Delta\chi^2 = 2$, well within the 90\% confidence range and a best-fit value for $\Gamma_2=3.2$, consistent with the results of the single power-law scenario. In conclusion, regions NW and SW represented two extreme possible regimes for jitter radiation, showing absence of  hard (jitter) and soft (synchrotron) components, respectively.  

 Since in the radio band a spectral index of $\sim 0.8$ is observed, we expected to measure $\Gamma_1 \sim 1.8$, or, by taking into account cooling break effects and the subsequent steepening $\Delta\Gamma = 0.5$ (K13), $\Gamma_1 \sim 2.3$. Therefore, we repeated the analysis described above by setting priors on the upper limit of $\Gamma_1$, i.e. leaving $\Gamma_1$ free to vary but forced to be lower than 2.5, leaving some space for spatial variation of the radio index across Cas A. We found energy break values lower than 4 keV, i.e. outside of the X-ray domain considered, and we obtained an overall worse description of the spectra for all the regions but NE and SE. However, a significantly ($> 3\sigma$) worse description was detected only for regions N ($\Delta\chi^2 \sim 30$), W ($\Delta\chi^2 \sim 70$) and Whole ($\Delta\chi^2 \sim 200$). Overall, these results indicated that: i) in regions N, W and Whole, some additional mechanism must be responsible for the radio-to-X-ray steepening of the synchrotron spectrum; ii) in the other regions, the X-ray synchrotron spectrum is compatible (within 3$\sigma$) with being the natural extrapolation of the radio one; iii) given the low value of $\ebreak$ outside of the X-ray domain considered, the broadband 4-40 keV spectrum is dominated by the jitter component in all regions but NE and SE. We will discuss in more detail this point in Sect. \ref{sect:disc}. 
 
\section{Discussion}
\label{sect:disc} 

We here reported on a spectral analysis of the Chandra/ACIS-S, NuSTAR/FPMA,B, SWIFT/BAT and INTEGRAL/ISGRI data of the SNR Cas A, demonstrating that the nonthermal emission 
is better described by the jitter model than by a model of a power-law with an exponential cutoff. In fact, in all regions but NW the 9-40 keV spectra are not well fitted at all by a simple cutoff.
The latter result was particularly evident for the whole remnant 
for which X-ray emission extends up to at least 100 keV, in disagreement with what we expect from a synchrotron regime beyond the cutoff. 
Additionally, there is supporting evidence at the $4\sigma$ level for nonthermal emission above 100~keV from CGRO/OSSE observations \citep{tlk96}.

The absence of a cutoff in the hard X-ray spectra is 
indicative of the presence of a radiation mechanism different from the standard synchrotron radiation. The jitter radiation model provides a well-argued and physically motivated alternative radiation mechanism for the origin of the nonthermal hard X-ray radiation. 
The photon spectrum of jitter radiation self-consistently takes into account the effect of the magnetic-field turbulence and directly connects it to the shape of the hard X-ray emission. In contrast, the standard model of a power-law spectrum with some form of exponential cutoff assumes an homogeneous magnetic-field, whereas, paradoxically, the high cutoff energy itself requires a highly turbulent magnetic field.
In the following sections, we discuss in detail the implication of these results on the turbulent magnetic field in which the electrons are embedded.

\subsection{Synchrotron versus jitter regimes}
The jitter radiation model (TF87, K13) is an extension of synchrotron radiation in presence of an highly turbulent magnetic field. For synchrotron radiation the acceleration is perpendicular to the magnetic-field causing the electrons to follow a helical path, with the magnetic field assumed to be constant at length scales comparable to the gyroradius of the electrons. Jitter radiation is a consequence of magnetic-field fluctuations on scales smaller than the photon formation length, causing deviations from the regular, helical path. 

At frequencies below $\omega_{\rm break}$ jitter radiation simply reduces to classical synchrotron radiation which dominates the radiative spectrum from the populations of relativistic electrons. If the electron energy distribution is a power law in energy---$n(E)\propto E^{-\xi}$, then the synchrotron radio spectral index is $\alpha=(\xi-1)/2$, translating into a photon index $\Gamma=\alpha+1$. For Cas A $\alpha=0.77$ and $\Gamma=1.77$, although a flattening of the average radio-to-infrared nonthermal spectrum
is likely happening above 10~GHz \citep[see the discussion in ][]{dmv21}. 

At the shock front the electron distribution is expected to have an exponential cutoff, caused by the fact that energy gains due to DSA are at high energies limited by radiative losses ---in the socalled loss-limited scenario---or alternatively by a limited time available for acceleration, the  time-limited scenario \citep{r96,hvb12}.
Further downstream of the shock, the maximum energy of electrons is lower as after having participated in the DSA process electrons keep losing energy.  Given the limited life time of a SNR, there is a maximum energy for
the electrons that were accelerated at the beginning. This gives rise to a break---the cooling, or age break---in the synchrotron spectrum with steepening of $\Delta \alpha=\Delta \Gamma=0.5$ \citep[e.g.][]{lon11}.
For Cas A the spectrum beyond the cooling break is expected to have an index $\alpha\approx 1.1$---1.3, or $\Gamma=2.1$--2.3.
This is clearly flatter than the power-law of the hard X-ray radiation, which is closer to $\Gamma\approx 3$ (see Table~\ref{tab:param_freeebreak}). 
Moreover, the nonthermal X-ray emission is clearly associated with narrow filaments at the shock front \citep{vl03} for which we expect the synchrotron spectrum to have an exponential cutoff, rather than a broken powerlaw.
The cutoff photon energy and the cutoff electron energy are related by the following:
\begin{equation}
\label{eq:syn_cutoff}
\hbar\omega_{\rm cut}\approx 0.19(B/100~{\rm \mu G})(E_{\rm e,cut}/10~{\rm TeV})^2~{\rm keV},
\end{equation}

supporting the idea of a cutoff photon energy in the range of 0.2--10 keV, given the uncertainty in $B$ and $E_{\rm e,cut}$.

Based on the X-ray synchrotron model of the filaments, the measured photon cutoff energy corresponds to an electron cutoff energy of around 10~TeV \citep{vl03}.
This is also in agreement with the cutoff energy inferred from the very-high energy gamma-ray emission measured by MAGIC \citep{aaa17m}.
Although, surprisingly, the cutoff inferred from the gamma-ray spectrum pertains to the hadronic (proton) cosmic-ray cutoff, rather than the
electron cutoff energy.
Clearly the lack of a clear cutoff energy in the hard X-ray spectra, the steepness of the spectra and the generally good fits provided by the \texttt{jitter} model lends support to the hypothesis that both the synchrotron and jitter components are contributing to the nonthermal X-ray emission from Cas A. An exceptional case is provided by the spectrum from the NW, where the \texttt{jitter} model suggests an unusually high break energy of $E_{\rm break}\sim 40$~keV, compared to 4--5 keV elsewhere.

\subsection{Magnetic-field turbulence spectrum and length scales}
\label{sect:turbulence_shape}

In this Section we take advantage of the results of the spectral analysis to infer significant information on the turbulence energy distribution and spatial scale.  Magnetic-field turbulence is an integral part of both collisionless shock heating and particle acceleration through the DSA mechanism. For DSA magnetic-field turbulence is needed at the scales of the Larmor radius. For the highest energy electrons this corresponds to $R_{\rm L}=E_{\rm e}/eB\approx 3\times 10^{14}(E/10~{\rm TeV})(B/100~{\rm \mu G})^{-1}$~cm. This is much larger than the turbulence length scales relevant for jitter radiation, which are below the non- or mildly relativistic Larmor radius, $R_{\rm L,j}\equiv m_{\rm e}c^2/eB=170(B/100~{\rm \mu G})$ km (K13).
This scale is therefore not so much relevant for the acceleration of relativistic particles through DSA, but it is for the plasma heating of collisionless shocks (and for the initial stages of particle acceleration), taking place on the ion inertial length scale, $\lambda_{\rm ii}= c/\omega_{\rm pi}\approx 230 n_{\rm p}^{-1/2}$~km. 

For synchrotron radiation the photon index $\Gamma_1 = \frac{\xi + 1}{2}$ reflects the spectral index $\xi$ of the electron spectrum. In contrast, in the jitter regime $\Gamma_2$ reflects the  spectral index $\nu_{\rm B}$ of the magnetic-field fluctuation spectrum (defined as in Eq. 69 of K13, with the relation $\Gamma_2=\nu_{\rm B}+1$. The spectral indices reported here (Table~\ref{tab:param_freeebreak}) range from  $\Gamma_2=3$  (NE) to 3.4 (N), implying $\nu_{\rm B}\approx 2.0$--   2.5. Models of magnetic-field fluctuation spectra that are often considered are $\nu_{\rm B}=5/3$ for Kolmogorov turbulence
 \citep{kol41} and $\nu_{\rm B}=3/2$ Kraichnan turbulence \citep{kra65}. Both are flatter than what is implied by the jitter model. However, Kolmorogov and Kraichnan turbulence assume injection of large scale turbulence
 at a large length scale cascading down to smaller length scales. At shock fronts the generation of turbulence is different. For example, \citet{tak23} argues that the Weibel instability results in a saturated spectrum with  
 $\nu_{\rm B}=2$. This would correspond to $\Gamma_2=3$, compatible with what we found here, and in line with the hard X-ray spectra of other young SNRs, which are also generally also around $\Gamma=3$ \citep[e.g. for SN1006,][]{lbm18}.

For synchrotron radiation the cutoff photon energy relates to the highest electron energy, see Eq.~\ref{eq:syn_cutoff}. For jitter radiation the maximum photon energy $E_{\rm{max}}$ is (K13)

\begin{equation}
E_{\rm{max}} \sim (R_{\rm{L,j}}/\lambda)^3\hbar\omega_ {\rm{cut}} \Rightarrow \lambda/R_{\rm{L,j}} = \sqrt{E_{\rm{break}}/E_{\rm{max}}},
\label{eq:omegamax}
\end{equation} 
for which we take advantage of the relation (K13)

\begin{equation}
\label{eq:break_synchtojitter}    
E_{\rm{break}}\sim \hbar\omegacut \frac{R_{\rm{L,j}}}{\lambda}, 
\end{equation}

with 
$\lambda$ the turbulence scale and

\begin{equation}
\omegacut=\frac{3}{2}\frac{eB}{mc}\Big(\frac{E_{\rm{e,cut}}}{mc^2}\Big)^2    
\label{eq:omegacut}
\end{equation}
being the cutoff frequency of the standard synchrotron regime (i.e. when R$_{\rm{L,j}} << \lambda$) with B the average magnetic field.

From Eq. \ref{eq:omegacut} we can resolve with respect to the cutoff energy of the population of electrons $E_{\rm{e,cut}}$  
emitting jitter radiation obtaining:

\begin{equation}
E_{\rm{e,cut}} = \sqrt{\frac{2}{3} \frac{\omegacut m^3c^5}{eB}} = \sqrt{\frac{2}{3} \frac{m^3c^5}{eB} \frac{E_{\rm{break}} \lambda}{\hbar R_{\rm{L,j}}}}
\label{eq:particles_cutoff}
\end{equation}
All the quantities in Eq. \ref{eq:particles_cutoff} are known, though we only have an upper limit on $\lambda$ and we then could put only upper limits on the particles' cutoff energy as well. The resulting values are shown in Table \ref{tab:turbulence}. 

For Cas A the downstream magnetic-field strength has been estimated to be in the range of 100--$500~{\rm \mu G}$ \citep{vl03,bv04,hvb12}.
For that reason we normalize to $B_{100}\equiv B=100~{\mu G}$, so that we can translate the best fit values of $ E_{\rm break}$ and our lower limits for $E_{\rm max}$ into an {\em upper limit} on the  minimum scale for the magnetic-field fluctuation length scale, $\lambda_{\rm min}$, as shown in Table~\ref{tab:turbulence}. For calculating $E_{\rm e,cut}$ we included an additional factor 2/3 in Eq.~(\ref{eq:syn_cutoff}), as suggested by K13.

The table shows that the condition \ref{eq:lambda} required for jitter to be at work is satisfied for all the regions but NW, indicating that magnetic-field fluctuations must be present to scales of $\sim 100$~km and smaller. Such values are remarkably small, and at a scale barely probed by in situ measurements of interplanetary shocks, which measure magnetic-field fluctuations with a resolution of seconds to minutes, which translates into length scales of 100--6000~km, for plasma speeds of 100~km/s. Noticeably, the cutoff energy of the particles in the jitter scenario does not differ much from the standard synchrotron, which predicts $E_{\rm e,cut} \sim 10 $ TeV.

\begin{table*}																
\centering																
\caption{Upper limits on the minimum turbulence scale and particles' cutoff energy.}													
\begin{tabular}{lccccc}\noalign{\smallskip}													
Region	&	$E_{\rm break}$	&	$E_{\rm max}$ (LL)	&	$\lambda/B_{100}$ (UL)	&	$E_{\rm ph,cut}$(UL)	&	$E_{\rm e,cut} B_{100}^{-1/2}$ (UL)			\\ 			
	&	(keV)	&	(keV)	&	(km)	&	(keV)	&	(TeV)			\\\noalign{\smallskip}\hline\noalign{\smallskip}			
Inner	&	4.5	&	40	&	57	&	1.5	&	15			\\			
North	&	5.2	&	40	&	61	&	1.9	&	17			\\			
Northeast	&	5.2	&	40	&	61	&	1.9	&	17			\\			
Northwest	&	40.4	&	40	&	171	&	41	&	78			\\			
SouthEast	&	6.0	&	40	&	66	&	2.3	&	19			\\			
Southwest	&	4.4	&	40	&	56	&	1.5	&	15			\\			
West	&	4.7	&	40	&	59	&	1.6	&	16			\\			
Whole	&	4.5	&	100	&	36	&	0.9	&	12			\\\noalign{\smallskip}\hline\noalign{\smallskip}			
\end{tabular}		

\textbf{Notes.} Scaled by $B_{100}=B/100~{\rm \mu G}$. LL and UL stand for lower and upper limit, respectively.
\label{tab:turbulence}
\end{table*}

\subsection{Polarization and filaments' width}
If the nonthermal hard X-ray emission from Cas A and other young SNRs is indeed due to jitter radiation then we need to rethink about some of the phenomenology of nonthermal X-ray emission that is often taken for granted. In particular, there are two aspects for which jitter radiation may have important consequences: X-ray polarization, and  the widths of X-ray synchrotron filaments.

So far two X-ray polarization measurements by IXPE of young SNRs have been reported, for Cas A \citep{vpf22} and Tycho's SNR \citep{fsp23}.
For Tycho's SNR the X-ray rim was reported to have 9\% polarized
X-ray emission, whereas for Cas A the polarization fraction was even lower: no isolated regions with a large polarization fraction could be found, and the overall polarization fraction on the nonthermal emission must be below 4\%, with a polarization angle topology consistent with radially aligned magnetic fields.
Again this argues for a highly turbulent magnetic fields at scales $\lesssim10^{17}$~cm in the context of synchrotron radiation. However, it is surprising that there is not a large scale anisotropy in the magnetic field, due to plasma shear---incurring radially oriented magnetic fields---or to shock compression causing tangential magnetic fields. Since jitter radiation is sensitive to very small-scale magnetic-field fluctuations, which are largely isotropic, it should not be intrinsically polarized.\footnote{Private communication Brian Reville. Note that \citet{pka16} calculated the expected polarization of jitter radiation, but assuming only 2 dimensional fluctuations, and showed that an anisotropic turbulence could still lead to a considerable amount of polarized light.} 
In the context of Cas A (and to a lesser extent Tycho's SNR), the low X-ray polarization fraction is simply a feature of jitter radiation. The small residual X-ray polarization suggest that a fraction of the non thermal radiation is ascribable to the synchrotron regime, and hence the synchrotron cutoff energy $E_{\rm ph,cut}$ should be in the soft X-ray band, as indeed seems to be the case according to Table~\ref{tab:param_freeebreak}.
A consequence that is potentially measurable should be that at larger photon energies---i.e. moving away from the synchrotron part of the spectrum---the polarization fraction should decrease. This is opposite to what is expected for pure synchrotron radiation, where at higher energies the spectrum steepens and originates from smaller regions, giving rise to a polarization fraction increasing with energy \citep[see][]{beo11}.

As for the narrow width of the nonthermal X-ray  filaments: these are usually interpreted as due to strong synchrotron losses of the high-energy electrons, giving rise to a width of $l_{\rm syn}=\frac{1}{4}V_{\rm s}\tau_{\rm syn}$,
with $V_{\rm s}$ the shock velocity and $\tau_{\rm syn}\approx 624/B^2E_{\rm e}$~s, the synchrotron loss timescale, or they are thought to be indicative of the electron diffusion length scale, 
$l_{\rm diff}\approx 4D_2/V_{s}$, with $D_2=\eta cE/eB$ the diffusion coefficient and $\eta=1$ in the Bohm limit. We see that for  $l_{\rm syn}$ the synchrotron filaments width should decrease with electron energy, and increase for $l_{\rm diff}$. 
However, near the cutoff electron energy one expects $l_{\rm syn}\approx l_{\rm diff}$ \citep[e.g.][]{vin20}.

Nevertheless, there have been attempts to measure the filament widths as a  function of photon energy. Most recently this was done by \citet{pwa23}, who reported that there is a " narrowing with energy of the synchrotron filaments in Cassiopeia A", and that the "energy dependency of this narrowing seems stronger at high energy \citep{pwa23}. In light of the synchrotron model this is somewhat surprising, as it suggest that filaments widths are purely defined by $l_{\rm syn}$ and not $l_{\rm diff}$.
However, these measurements can be explained by jitter radiation, considering that the small-scale magnetic-field fluctuations generated near the shock dampen further downstream of the shock. This argument is somewhat
reminiscent of the work of \citet{pyl05}, but that argued for a decay of the overall magnetic field, rather than a decay of the magnetic-field turbulence.
So jitter radiation offers a natural explanation for the narrowing of the filament widths with energies. In the context of Cas A it would imply again that at softer X-ray energies the width is partially determined by the
combination of $l_{\rm diff}$ and $l_{\rm syn}$, but at higher energies, where the spectrum becomes more dominated by the jitter component, the width starts to be determined by the length scale over which small-scale
turbulence is dampened. In principle this suggest that the narrowing of the filaments should only become prominent for energies beyond $E_{\rm break}$, consistent with the findings of \citet{pwa23}.

\section{Summary and conclusions}
\label{sect:conc}

In this paper we presented our results on the analysis of multi-instrument X-ray data of the SNR Cas A aiming at characterizing the shape of nonthermal emission in a wide energy band and in investigating its origin. We analyzed Chandra/ACIS-S, NuSTAR/FPMA,B, SWIFT/BAT and INTEGRAL/ISGRI observations of Cas A in the 4-100 keV energy band and performed a spatially resolved spectral analysis adopting various nonthermal spectral models. Our findings can be summarized a follows.

\begin{enumerate}
    \item While DSA intrinsically requires high magnetic turbulence, the spectral models used to fit nonthermal spectra of SNRs do not include the effect of turbulence in the shaping of the spectra. The standard approach to fit SNRs spectra in the 4-6 keV energy band to investigate its synchrotron radiation is insensitive to the actual shape of the nonthermal model adopted.
    \item The 15-100 keV spectra of the whole Cas A can be adequately fitted only through a power-law with a photon index $\Gamma \sim 3.3$. The 15-100 keV spectra are not well fitted either by the \texttt{srcut} \citep{rk99} or the \texttt{zira}\citep{za07} model, as there is no evidence for an exponential cutoff.
    \item A jitter model describes the 4-40 keV spectra of all the regions considered much better than any other model showing a clear steepening of the spectrum non-compatible with a cutoff and with a best-fit photon index $\Gamma_2 \sim 3-3.4$.
    
    \item The best-fit photon index $\Gamma_2 \sim 3-3.4$ imply a spectral index for the turbulence spectrum $\nu_B \sim 2-2.4$, higher than common values such as Kolmogorov $\nu_{\rm{Kol}}=5/3$ and Kraichnan $\nu_{\rm{Kra}}=3/2$. This applies to magnetic-field fluctuations on scales of $\sim 100$~km, much smaller than the scales normally invoked for the DSA mechanism, and more in line with fluctuations near collionless shocks as induced by the Weibel instability.
    
    \item We estimated upper limits on the minimum turbulence scale
    which are typically smaller than  $100$~km, whereas the estimated transition
    energies from synchrotron to jitter regimes imply cutoff electron energies of typically 15~TeV, as in the standard synchrotron scenario.

    \item If indeed the nonthermal X-ray emission is due to both synchrotron and jitter components this offers an alternative natural explanation for the low polarization fraction for Cas A and the narrowing of nonthermal X-ray filaments with increasing energy. So, potentially, with nonthermal X-ray emission we are zooming into regions very close to
    the collisionless shocks themselves. 
\end{enumerate}

\bibliography{references,new}
\bibliographystyle{aasjournal}

\section*{Acknowledgements}
This project has received funding from the European Union’s Horizon 2020 research and innovation program under grant agreement No. 101004131 (SHARP). We thank the XSPEC helpdesk and R. Kristin for the help in the implementing of the jitter model.
We thank Brian Reville for helpful comments on jitter radiation. 

\begin{appendix}
\section{Zira and jitter models in XSPEC}
\label{sect:jitter_xspec}

\subsection{Zira}

We implemented the Eq. 37 in \citet{za07} through the \texttt{mdefine} task available within XSPEC, defining a new \texttt{zira} model. The corresponding equation in the notation used in this paper can be read as:

\begin{equation}
(E_{\rm{rad}}/E)^2(1+0.38\sqrt{E/\ecut})^{11/4}exp[-(E/\ecut)^{1/2}]    
\end{equation}

where $E_{\rm{rad}}$ is always kept fixed to 1. The exponential 2 in $(E_{\rm{rad}}/E)^2$ is added to match the default XSPEC units. 
\subsection{Jitter}

The total \texttt{jitter} model does not have an analytical expression and cannot be included within XSPEC through the \texttt{mdefine} task. We followed the approach described in \href{https://heasarc.gsfc.nasa.gov/xanadu/xspec/manual/XSappendixLocal.html}{Load\_model\_XSPEC} to include our \texttt{jitter} model in the library of the available models. The synchrotron dominated and jitter dominated regimes are respectively modeled by two power laws with an additional exponential cutoff for the low energy component. However the simple addition of both components in a single model does not replicate the complex shape of the total emission, and advocates for additional energy and/or flux dependent constraints. For this purpose the flux given by the two components is computed iteratively in each energy bin (starting from low energy) and the highest is kept as the model flux. The ratio of both component flux, labeled \texttt{jratio}, is added as a parameter to the model and acts as a normalization. For continuity and self-consistency reasons, the jitter flux is set to 0 for energies lower than the synchrotron break, preventing any ill behavior in situations of steeper jitter slope resulting in high flux at low energies. We tested our model by comparing it with other well-known models such as \texttt{pow}, \texttt{bknpower}, i.e. a broken power-law, and \texttt{cutoffpl}, a power-law with an exponential cutoff, in regimes in which the jitter radiation simply reduces to one of these.

The \texttt{jitter} model is finally described by six parameters: \texttt{PhoIndex1}: synchrotron power-law slope ($\Gamma_1$), \texttt{Ebreak1}:  synchrotron break energy ($\ebreak=\hbar\omega_{break}$, with $\omega_{break}$ defined in Fig. \ref{fig:toptygyn}), \texttt{PhoIndex2}: jitter-radiation power-law slope ($\Gamma_2$), \texttt{jratio}: ratio between flux in regime 1 and regime 2 ($P_2/P_1$ in TF87), \texttt{beta1}: shape of the exponential cutoff around \texttt{Ebreak1} ($\propto \exp[- (\omega/\omega_{break})^{\beta}]$), \texttt{norm}: total normalization of the model.

\section{Cross-calibration between X-ray telescopes}
\label{app:cross_calibration}
We investigated potential issues causing the significant discrepancy observed in the Chandra and NuSTAR spectra in the 4-6 keV energy range. The distortion in the Chandra/ACIS-S spectra could be produced by pileup effects. We produced, through the \texttt{pileup\_map} task within {\it CIAO}, a pileup map of Cas A, finding that the average pileup fraction in the regions considered is of roughly 3\%, and it was therefore excluded as a possible origin for the flattening. We checked that the flux of NuSTAR spectra in the $>$15 keV energy band is consistent with those reported for BeppoSAX  (\citealt{vl03}) and INTEGRAL (\citealt{rvd06}) considering the nominal cross-calibration constant (\citealt{mhm15,mfg22}). We used \texttt{srcflux} on the Chandra image in the 4-6 keV band and checked that the observed counts matched the counts in the same energy range in the spectra. We repeated this procedure also for the NuSTAR data with the ftool \texttt{countsinregion}. Finally, we used adopted different values for the parameter \texttt{boxsize}=5,10,20,30,40 obtaining identical spectra and set \texttt{flatflagarf=yes} as in \citet{gfh17}. 

\section{SWIFT upper limit on flux}
\label{sect:swift_upperlimit}
We estimated the upper limit on the X-ray flux of A Cas above 100 keV by estimating, through the \texttt{error} command within XSPEC, the 90\% and 99.7\% confidence level (corresponding to $\Delta\chi = 2.706$ and $\Delta\chi^2 = 9$) starting from the SWIFT/BAT spectra of Cas A. The best-fit model was a power-law having $\Gamma=3.5$. The error on the flux was estimated through the \texttt{cflux} component. The gaussian components, describing the lines of Ti44, are kept fixed since these are not resolved by SWIFT but contribute to the global flux. The corresponding upper limit on the flux between 100 and 200 keV were $0.28 \times 10^{-11}$ erg/s and $0.33 \times 10^{-11}$ erg/s at the 90\% and 99.7\% confidence levels, respectively.

\section{Nonthermal to total flux ratio}
\label{sect:nonthermalratio}

In Sect. \ref{sect:internal} we showed that, we cannot rule out or confirm the presence of a cutoff in the 4-6 keV spectra. As already mentioned, a possible source of uncertainty is the thermal emission from the shocked-heated plasma. Because of the bad spatial resolution of NuSTAR and the need for high statistics up to 40 keV, we selected regions with a characteristic size at least comparable to the NuSTAR PSF ($\sim 60''$ in Half Power Diameter). It was then natural to expect significant shocked ejecta signatures in the extracted spectra. This additional thermal contribution is added to the putative nonthermal cutoff that is therefore blurred out.

\begin{figure}[!ht]
    \centering
    \includegraphics[width=0.9\columnwidth]{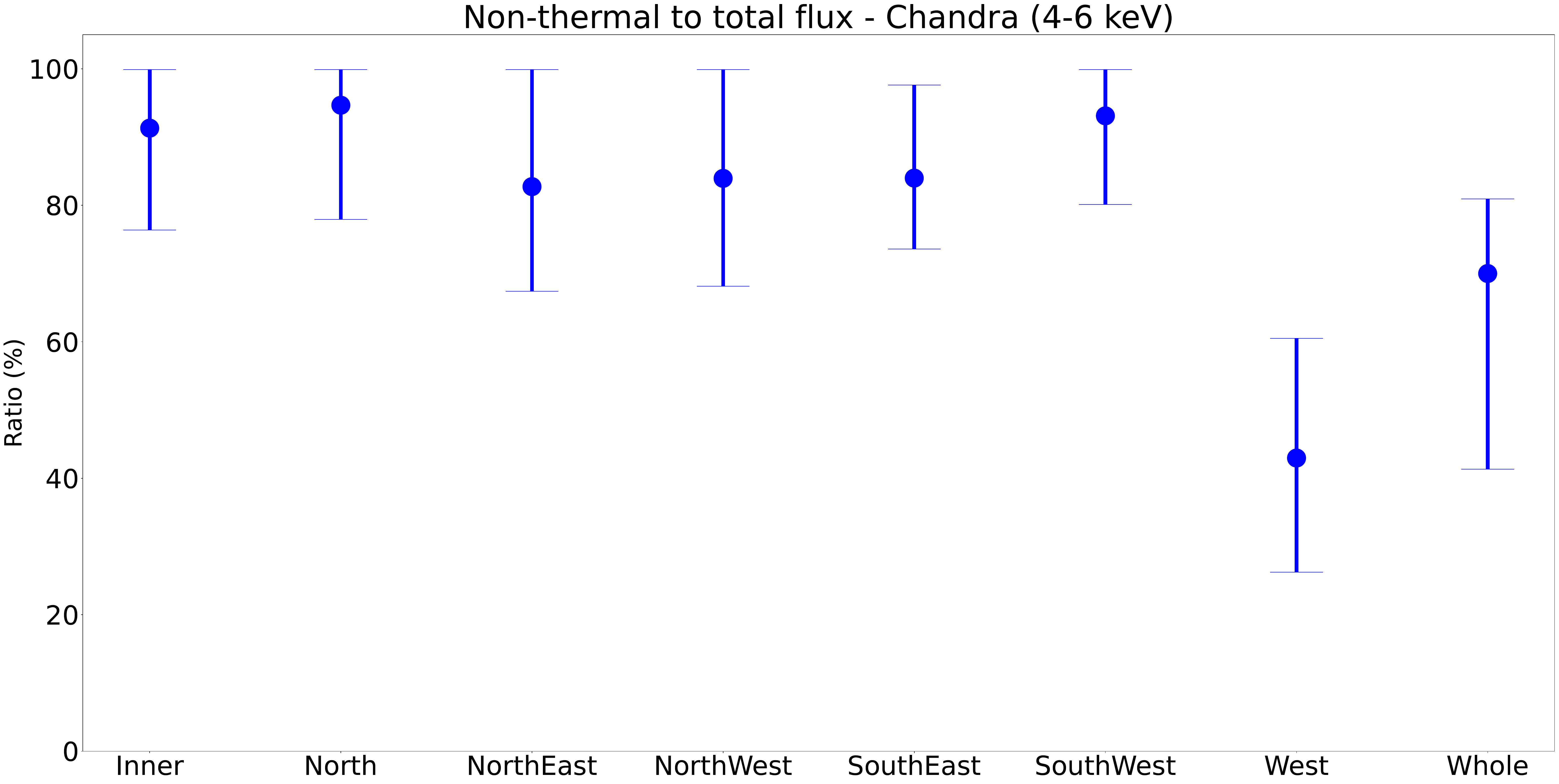}
    \caption{Ratio nonthermal over total flux in the 4-6 keV band obtained from the fits in the 0.5-8 keV band as a function of the region.}
    \label{fig:ratio}
\end{figure}

A partial, solution for this issue is to measure the nonthermal and the thermal flux from Chandra/ACIS-S best-fit models performed in the 0.5-8 keV band. We modeled the Chandra/ACIS-S data with a model made of an absorbing component, \texttt{TBabs} model in XSPEC, a power-law and a number of \texttt{vnei} components, depending on the complexity of the plasma thermal emission. Given the extremely high number of counts in the spectra extracted from regions I, W and Whole, we included a systematic error of 3\%, 3\% and 5\% respectively during the fitting procedure. We measured the total and the nonthermal only flux through the \texttt{cflux} component available within XSPEC. In Fig. \ref{fig:ratio} the ratios nonthermal to total flux for all the regions considered are shown.

All the regions but West and Whole showed a nonthermal to total flux ratio higher than 80\%, and consistent with 100\%. Lower values found for West and Whole regions were not surprising since in these regions significant emission from shocked ejecta was revealed in previous works (e.g. \citealt{hl12}). These results confirmed that not including a thermal component when fitting 4-6 keV energy range had a small to zero effect on the estimate of the slope of the spectra. It is worth to mention that the values displayed in Fig. \ref{fig:ratio} did not provide information on the shape of the photon spectra because of the degeneracy between emission measure and abundance (see \citealt{gvm20} for details). In fact, by repeating the analysis with pure-ejecta components, i.e. \texttt{vnei} models with abundances values fixed to $ \gtrsim 10^3$, the ratio increased by reaching values of 95$\%$, as also reported by \citet{vpf22}.

\newpage

\section{Plots for all the regions}
\label{app:two-bands}

\begin{figure*}[!ht]
    \centering   
    \includegraphics[height=.9\textheight,width=.9\textwidth]{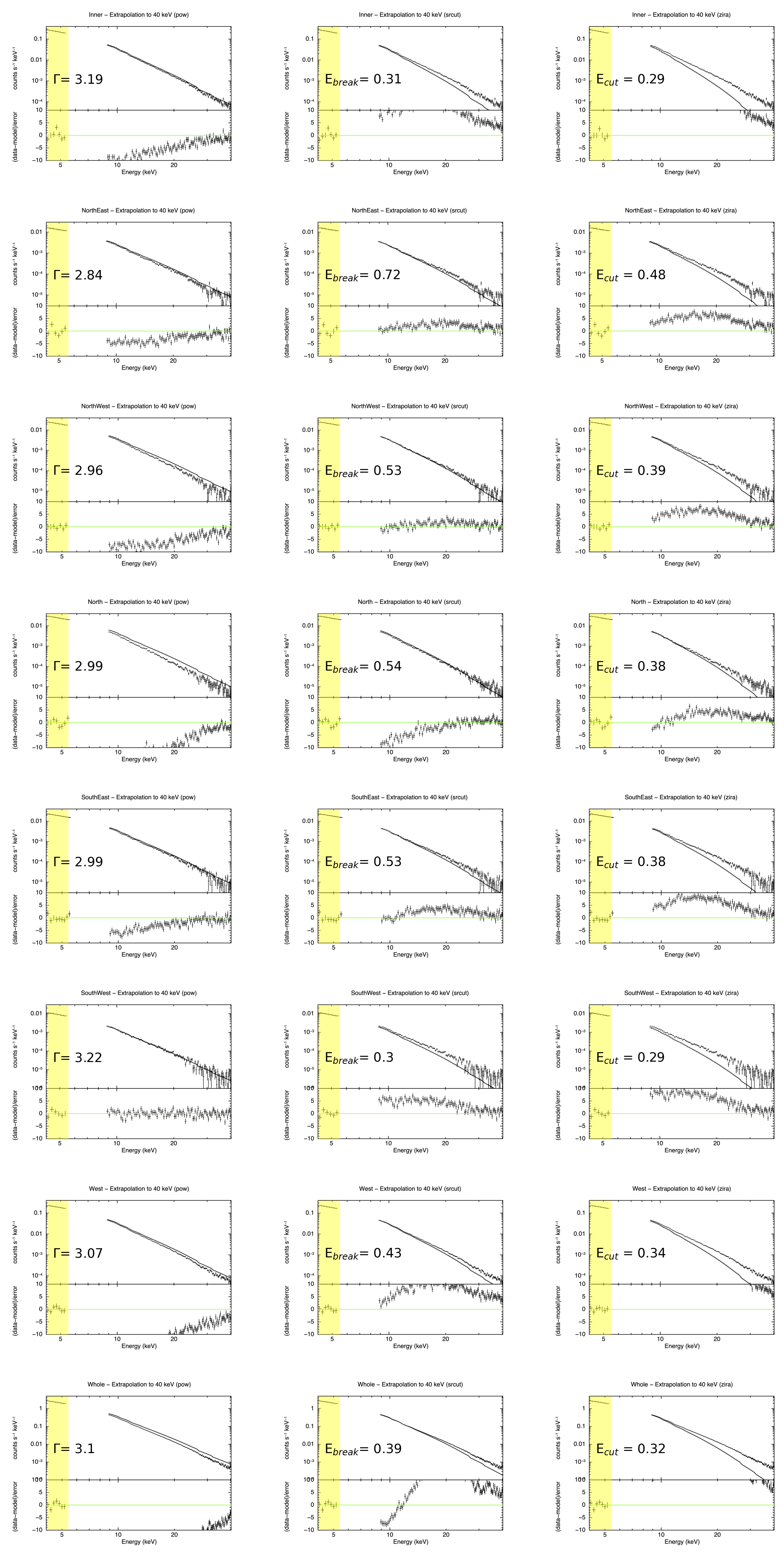}
    \caption{Same as Fig. \ref{fig:fixebreak_singleregion} but for all the regions considered}
    \label{fig:fixebreak_4-9-40}
\end{figure*}

\begin{figure*}
    \centering    
    \includegraphics[height=.9\textheight,width=.9\textwidth]{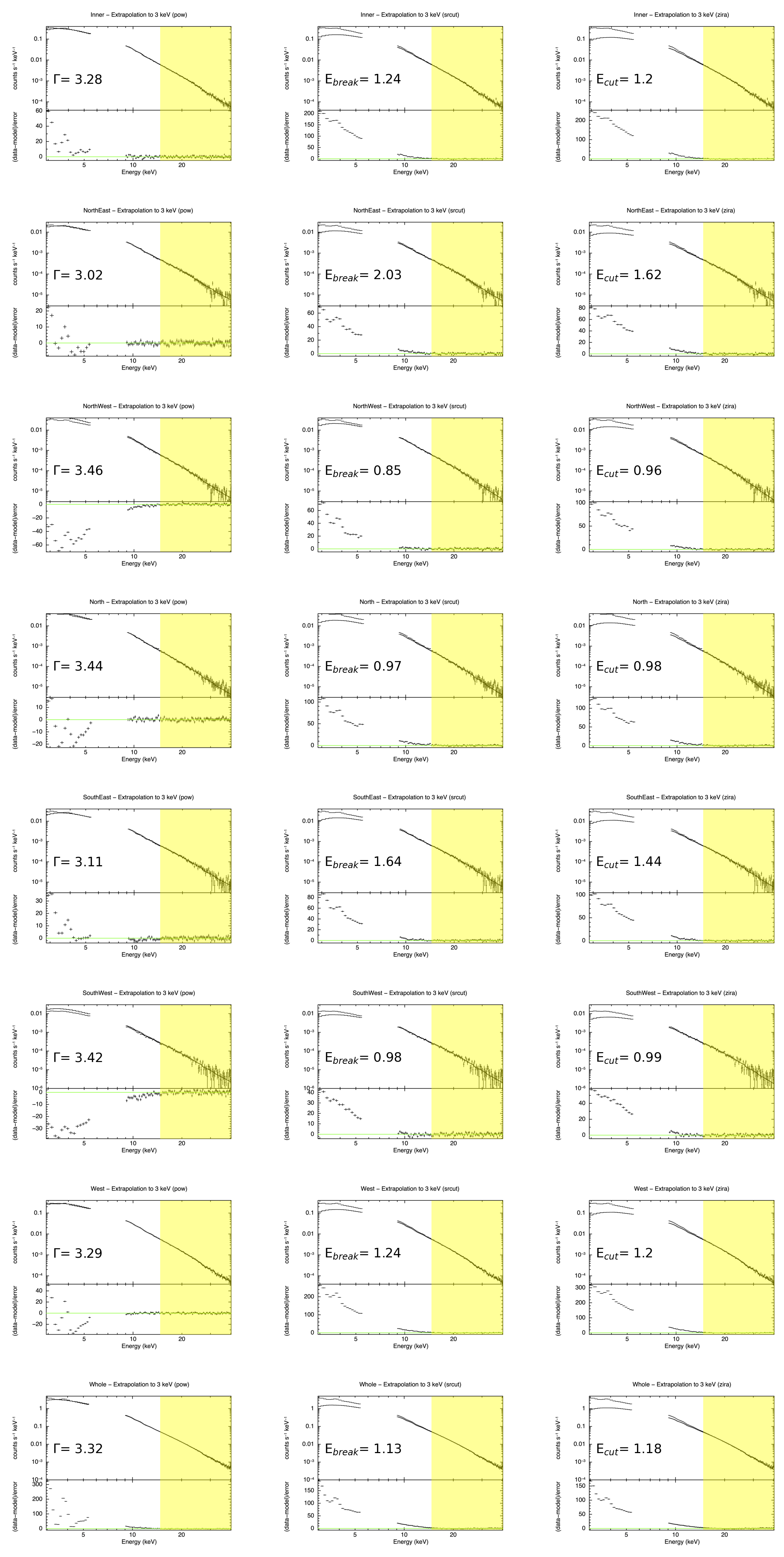}
    \caption{Same as Fig. \ref{fig:fixebreak_9-40_to_4-6_singleregion} but for all the regions}
    \label{fig:fixebreak_9-40_to_4-6}
\end{figure*}

\begin{figure*}[!ht]
    \centering
    \includegraphics[height=.93\textheight,width=.95\textwidth]{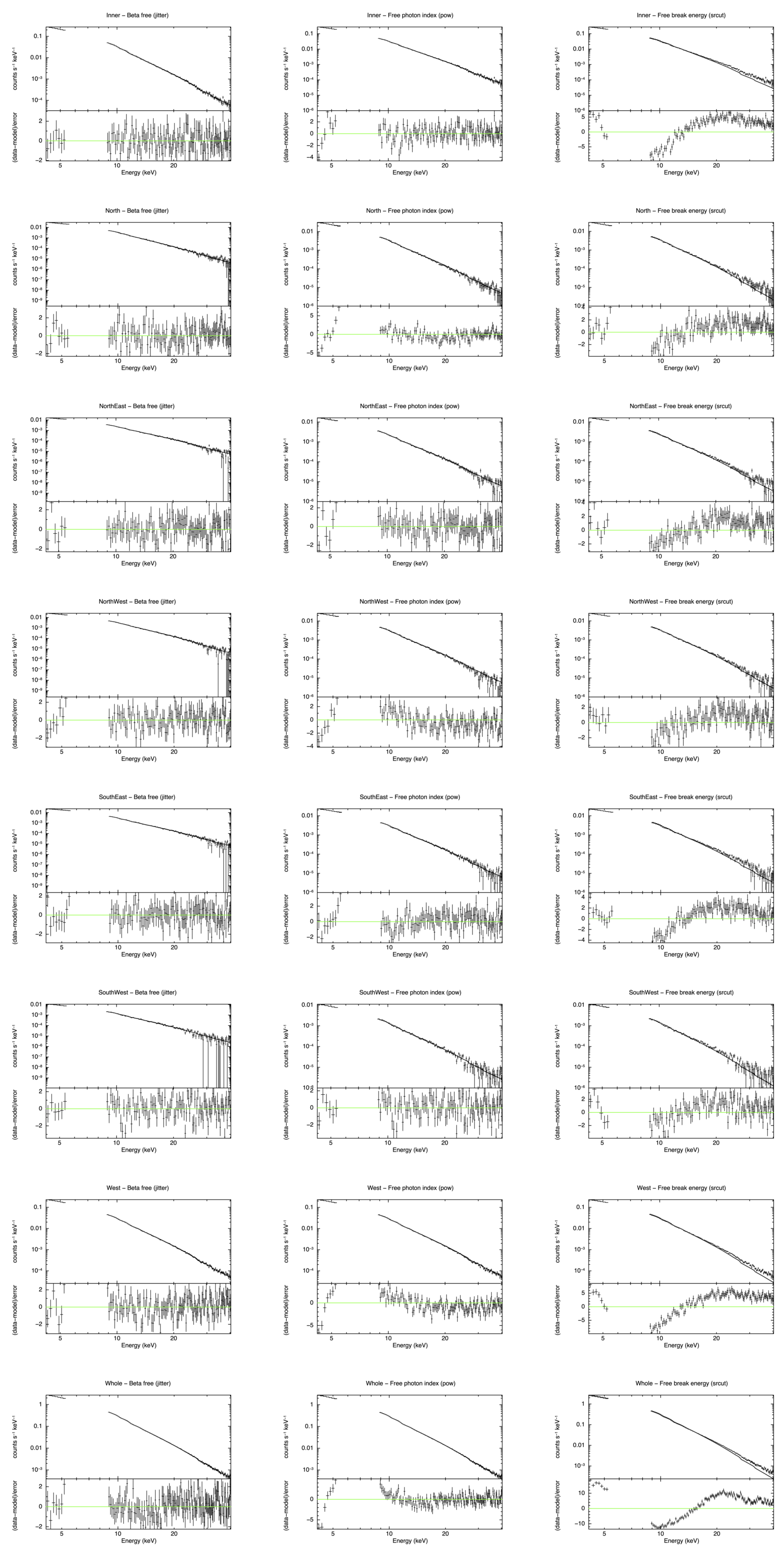}
    \caption{ Same as Fig. \ref{fig:freebreak_4-9-40_singleregion} but for all the regions}
    \label{fig:freebreak_4-9-40}
\end{figure*}

\end{appendix}

\end{document}